\documentclass[aps,prb,amsmath,twocolumn,amssymb,titlepage,superscriptaddress,showpacs]{revtex4-1}
\usepackage[english]{babel}
\usepackage{graphicx}
\usepackage{transparent}
\usepackage{amsmath}
\usepackage{amssymb}
\usepackage{epstopdf}
\usepackage{amsfonts}
\usepackage{bm}
\usepackage{mathdots}
\usepackage{times}
\usepackage{pdfpages}
\usepackage{mathtools}
\usepackage{stmaryrd}
\usepackage{tabularx}
\usepackage{hhline}

\makeatletter
\AtBeginDocument{\let\LS@rot\@undefined}
\makeatother

\newcommand{\ave}[1]{\langle #1 \rangle}

\newcommand{\beq}[1]{\begin{equation} #1 \end{equation}}

\newcommand{\astcycl}{\mathrlap{\kern0.085em{\circlearrowright}}\ast}
\newcommand{\taucycl}{\mathrlap{\kern0.42em{\bullet}}\circlearrowright}

\begin{document}
\title{A comparative study of nonequilibrium insulator-to-metal transitions in electron-phonon systems}
\author{Sharareh Sayyad}
\affiliation{Institute for Solid State Physics, University of Tokyo, Kashiwanoha,
Kashiwa, 277-8581 Chiba, Japan}
\author{Rok \v Zitko}
\affiliation{Faculty of Mathematics and Physics, University of Ljubljana,
Jadranska 19, SI-1000 Ljubljana, Slovenia
}
\affiliation{
Jozef Stefan Institute, Jamova 39, SI-1000 Ljubljana, Slovenia
}
\author{Hugo U.~R.~Strand}
\affiliation{Center for Computational Quantum Physics, Flatiron Institute, 162 Fifth Avenue, New York, NY 10010, USA}
\author{Philipp Werner}
\affiliation{Department of Physics, University of Fribourg, 1700 Fribourg, Switzerland}
\author{Denis Gole\v z}
\affiliation{Department of Physics, University of Fribourg, 1700 Fribourg, Switzerland}


\begin{abstract}
We study equilibrium and nonequilibrium properties of electron-phonon systems described 
by the Hubbard-Holstein model using the dynamical mean-field theory.
In equilibrium, we benchmark the results for impurity solvers based on
the one-crossing approximation 
and 
slave-rotor approximation
against non-perturbative numerical renormalization group reference data.
We also examine how well the low energy properties of the
electron-boson coupled systems
can be reproduced by an effective static electron-electron interaction. 
The one-crossing and slave-rotor approximations are then used to
simulate insulator-to-metal transitions induced by a sudden switch-on of the electron-phonon interaction. 
The slave-rotor results suggest the existence of a  critical electron-phonon
coupling above which the system is transiently 
trapped in a non-thermal metallic
state with coherent quasiparticles. The same quench protocol in the one-crossing approximation results in a bad metallic state.
\end{abstract}
\maketitle

\section{Introduction}

A Mott insulator can be realized in correlated lattice systems
if the interaction energy is comparable to or larger than the kinetic energy.
In such systems, changes in thermodynamic parameters may induce insulator-to-metal transitions (IMTs),  
as has been demonstrated by varying temperature~\cite{Morin1959} or pressure.~\cite{Qiu2017,Gavriliuk2012}  
Laser-induced mechanisms provide another strategy to manipulate
quantum phases in these materials.~\cite{Perfetti2006,Tobey2008,Hu2016,Kaiser2017} In these experiments,
phase transitions or transitions to metastable states are induced by a time-dependent perturbation. 
The resulting dynamics often follows a highly non-thermal trajectory and in the context of IMTs 
interesting questions arise concerning both the timescale and the pathway for the non-adiabatic switching. 

The essence of the correlation-induced IMT is encapsulated in the
Hubbard model.~\cite{Eckstein2013,Sayyad2016} As the initial Mott 
insulating phase has a large repulsive electron-electron interaction
which localizes electrons, a transition to a metallic state 
can be achieved by enhancing the screening originating
either from the coupling to lattice degrees of
freedom~\cite{Werner2007,Koller2004} or plasmonic
excitations.~\cite{Werner2010,Golez2015}
The theoretical description of these processes involves extensions of the
Hubbard model which incorporate the effect of 
electron-phonon~\cite{Jeon2004,Sangiovanni2005,Sangiovanni2006} or
nonlocal Coulomb interactions.~\cite{Golez2015,Kapcia2017,Schuler2017,Gao2009}
The proper description of screening effects is particularly important due to the
large change in the number of mobile charge carriers during the excitation and IMT.

In this work, we will focus on IMTs triggered by a time-dependent change 
in the strength of the electron-phonon coupling. The later can be realized by 
terahertz driving and it is enhanced via anharmonic effects.\cite{babadi2017,murakami2017,kennes2017,sentef2017} 
We will consider the Hubbard-Holstein model, where the electrons interact through an on-site
Coulomb repulsion and are linearly coupled to 
dispersionless phonons. The equilibrium phase diagram of the
Hubbard-Holstein model contains metallic and Mott-insulating phases
as well as a bi-polaronic insulating phase.~\cite{Jeon2004,Werner2007}
Equilibrium studies of the Hubbard-Holstein model have revealed that the dynamical nature of the phonon-induced
effective electron-electron interaction cannot be neglected, except in the 
large-phonon frequency limit,~\cite{Werner2013} and it is responsible for
the different behavior in the high- and low-energy regimes.~\cite{Sangiovanni2005, Sangiovanni2006}
Despite a strong influence of the phonons on the high-energy part of the spectrum, the
low-energy physics can be described by the Hubbard model with an appropriately determined 
reduced static interaction. In this study, we consider time-dependent modulations
of this screened interaction and the resulting IMTs. We will be interested in a quantitative
description of the nonequilibrium transition into the metallic phase 
and the corresponding thermalization time. 

Simulating the nonequilibrium dynamics of a strongly
correlated system coupled to phononic degrees of freedom is a
challenging problem. In weakly coupled systems, phonons can
either be treated by the Migdal approximation with~\cite{Murakami2015,schuler2016}
or without~\cite{sentef2013,rameau2016,kemper2013}
a self-consistent renormalization of the phonon propagator.
In the former case the mutual interaction between the electronic and phononic
subsystems self-consistently screens the static Coulomb
interaction and renormalizes the phonon energy. Strongly interacting
electron-phonon coupled systems have been studied within the dynamical mean field theory (DMFT)
approximation.~\cite{Werner2013,werner2015} In contrast to the
equilibrium case~\cite{Werner2007,Koller2004} powerful exact
solvers for non-equilibrium electron-phonon coupled impurity
problems are missing. It is thus important to benchmark and compare the existing state-of-the-art impurity solvers which
can be extended to non-equilibrium situations.
In this study, we focus on the one-crossing approximation\cite{grewe1981,coleman1984,eckstein2010} and slave-rotor\cite{Florens2002,Sayyad2016} based impurity solvers and
compare equilibrium spectra and phase diagrams against numerically exact
reference calculations obtained by the numerical renormalization group
(NRG).\cite{wilson1975,krishna1980a,bulla2008} This provides information about the parameter regimes in which the  approximate impurity solvers produce reliable results. In the second part of this work, we
compare the time evolution predicted by the approximate impurity solvers
and address the question of nonequilibrium IMTs.

The outline of this paper is as follows. In Sec.~\ref{sec:Ham} we
introduce the model Hamiltonian and the associated
dimensionless parameters. Section~\ref{sec:method} explains the three
approximate impurity solvers used in the paper.
In the first part of Sec.~\ref{Sec:eq} we show extensive comparisons of the equilibrium
spectral functions obtained from different approximations.
The second part is devoted to the study of
the low-energy properties of the correlated metal and
the search for a purely electronic Hamiltonian which effectively describes the low
energy physics. In Sec.~\ref{sec:neq} we present the time evolution
after a sudden quench of the electron-phonon coupling and
discuss the appearance of a non-thermal transient state with an enhanced quasi-particle weight
in the slave-rotor calculations. Section~\ref{sec:Con} contains a brief conclusion. 

\section{Model Hamiltonian}\label{sec:Ham}

The half-filled one-band Hubbard-Holstein model is described by the Hamiltonian
\begin{align}
  H_{\rm HH}= &
                -v \sum\limits_{\ave{ij}\sigma} c^{\dagger}_{i\sigma} c_{j\sigma}
                +U \sum\limits_{i} n_{i \uparrow} n_{i \downarrow} \nonumber \\
              &+\omega_{0} \sum\limits_{i} b^{\dagger}_{i} b_{i}
                +g \sum\limits_{i \sigma} (n_{i}-1)  \big( b^{\dagger}_{i}+b_{i} \big)  , 
 \label{eq:HamHH}
\end{align}
where $c^\dagger_{i \sigma}(c_{i \sigma})$ is the electron
creation~(annihilation) operator at site $i$ with spin $\sigma = \pm \frac{1}{2}$, $n_i=n_{i\downarrow}+n_{i\uparrow}$ and
$b^{\dagger}_{i}(b_{i})$ creates~(annihilates) a phonon at site $i$.
The first term of Eq.~\eqref{eq:HamHH} describes the hopping of an
electron with spin $\sigma$ from site $j$ to one of its
nearest-neighbors $i$ with amplitude $v$ that determines the
bandwidth $W$. The electrons locally interact with a Coulomb
repulsion $U$. This electronic system is coupled to Einstein phonons
with frequency $\omega_{0}$ by a linear coupling $g$ between the
local density of electrons and the phonon displacement.
Throughout this paper, we use a bandwidth $W=4v$ and set
$v~(1/v)$ as the unit of energy~(time).

The system is parametrized by three dimensionless parameters, namely
(i) the ratio between the electron-electron (el-el) interaction and
bandwidth $U/W$ which controls the insulating tendency of the system,
(ii) the dimensionless electron-phonon~(el-ph) coupling
$\lambda=g^2/v\omega_0$ which measures the gain of energy due to the
el-ph coupling in the atomic limit, and (iii) the adiabaticity
of the phonon $\omega_0/W$ which determines the relative speed of the
phononic and electronic degrees of freedom. The equilibrium phase
diagram~\cite{Werner2007,Koller2004} (excluding symmetry broken
phases \cite{murakami2013,murakami2014}) at half-filling results from a competition between these
effects. For weak el-ph coupling $\lambda \ll U/(2v)$ the
system exhibits a metal-to-insulator~(Mott) transition due to the
el-el interaction. The metallic and the Mott insulating
states are driven into a bi-polaronic insulating state by increasing
the coupling $\lambda$ to the bosonic degrees of freedom. However,
since phonons are coupled to charge fluctuations, which are strongly
suppressed in insulators, the correlation functions in metals are
expected to be more sensitive to the el-ph coupling than in
insulators. In the adiabatic limit, $\omega_0/W\ll 1$ the Migdal
theorem states that the vertex corrections are small and the transition to the bi-polaronic state will occur at intermediate
el-ph coupling $\lambda,$ while away from the adiabatic
limit the critical coupling increases.\cite{Werner2010} 
In Sec.~\ref{Sec:eq} we will demonstrate the characteristic behaviors in the various
regimes of the phase diagram by comparing the spectral 
functions obtained from different approaches in order to assess
the validity of the employed approximations in the parameter space.

\section{Numerical methods}\label{sec:method}

Our numerical investigation is based on the dynamical mean
field theory~(DMFT).~\cite{georges1996,Aoki2014} 
This approximate method assumes a spatially local self-energy and 
maps the lattice model onto a self-consistent solution of
a quantum impurity model which is coupled to a bath. 
The formalism becomes exact in the limit of infinite coordination number and our calculations with 
a semi-circular density of states corresponds to a Bethe
lattice in this limit.
The main limitation which determines the accuracy of the DMFT solution in this limit is the impurity solver. 
While in equilibrium powerful non-perturbative methods have been developed to
solve impurity problems coupled to bosonic degrees of freedom,
such as quantum Monte Carlo~\cite{Werner2007,Assaad2007} or numerical
renormalization group (NRG)\cite{bulla2008} solvers, there exists no
numerically exact and efficient approach to treat the nonequilibrium
situation. For this reason, several approximate nonequilibrium solvers have been
developed. Each has specific merits, applicability restrictions, and numerical demands.
In order to understand their
limitations we will collect DMFT
solutions obtained from three different impurity solvers. By comparing
the equilibrium spectral functions and by considering NRG results as an accurate reference point, 
we obtain insights into the features that are properly described, as
well as the range of validity of the different solvers.

In the following we briefly describe some relevant properties of the impurity solvers used in this work:

\begin{enumerate}

\item The strong-coupling perturbation method based on a 
self-consistent diagrammatic expansion in the hybridization
function, which at the first~(second) order is known as the
non-crossing~(one-crossing) approximation
NCA~(OCA),\cite{grewe1981,coleman1984,eckstein2010} has been
extended to el-ph interacting problems via an additional weak coupling
expansion in the el-ph coupling
strength.\cite{Golez2015,haule2003,chen2016} A detailed
description of this combined strong/weak coupling approach can be found in
Ref.~\onlinecite{Golez2015}. In the following, we will employ the
OCA approximation, since the description of the correlated metal is
significantly improved in comparison to the NCA counterpart. By construction, this
method is limited to strong el-el interactions
and to the weak el-ph coupling regime and will be referred to
as OCA-WC. For strong electron-phonon coupling a complementary approach\cite{Werner2007,Werner2013} can be formulated using the Lang-Firsov transformation.\cite{Lang:1963aa} However, since this 
approximation is not well-behaved in the small-$\omega_0$ limit, 
we defer the discussion of this method to Appendix \ref{Supp:LF}.

\item The slave-rotor~(SR) decomposition has been employed to solve the impurity problem
in Refs.~\onlinecite{Florens2002,Sayyad2016}.
In this work, we fix the fudge parameter of the SR to ${\cal N}=3$ to adjust the phase diagram at $g=0$.
In the self-consistent weak el-ph coupling approximation,
one can substitute the interacting phononic Green's function,
instead of the dissipative propagator, into the
slave-rotor method, see Ref.~\onlinecite{Sayyad2016}. 
To obtain the interacting phononic Green's function, we employ the updating procedure of the weak-coupling expansion described in Ref.~\onlinecite{Golez2015}. 
The combination of the weak el-ph coupling expansion
and the slave-rotor decomposition restricts this impurity solver~(denoted SR-WC)
to the physics in the weak el-ph coupling regime.

\item The numerical renormalization group~(NRG) method\cite{wilson1975,krishna1980a,bulla2008} can be easily extended to
incorporate local phonon modes by expanding the impurity basis with a
vibrational degree of freedom.\cite{hewson2002,meyer2002,jeon2003,Jeon2004,cornaglia2004}
This approach has found many applications in the context of quantum
transport through vibrating molecules and for bulk systems via the DMFT mapping.\cite{Koller:2004ic,koller2004euro,cornaglia2005,koller2005phonon,
cornaglia2007,Bauer:2010hx,Bauer:2010by,golez2012} 
The phonon cutoff needs to be increased until convergence is reached.
This implies that the calculations become numerically costly when the phonon
mode softens close to the transition into the polaronic state.
In this work, most calculations are 
performed with a phonon cutoff set at ten, with the NRG discretization
parameter $\Lambda=2$ (or $\Lambda=2.5$ for mapping out the phase
diagrams), keeping all multiplets up to an upper cutoff energy
$8$ (in units of the characteristic energy scale at the $N$-th step of
the iteration), with $N_z=4$ interleaved discretization grids.\cite{resolution,odesolv} 
To study finite temperatures, we made use
of the full-density-matrix algorithm.\cite{anders2005,peters2006,weichselbaum2007}

\end{enumerate}

In the SR-WC and OCA-WC approaches, we employ the non-equilibrium Keldysh
formalism to describe the time evolution.\cite{Aoki2014}
The spectral properties are obtained
by the real-time propagation of the solution and a partial Fourier transform. 
For example, the spectral function is obtained as
\beq{
A(\omega,t)=-\frac{1}{\pi}\text{Im}\int_0^{t_\text{max}} {\rm d}t' e^{\mathrm{i} \omega t' }G^{R}(t+t',t),
\label{Eq:Aw}
} 
where the typical value for the integration window is given by
$t_\text{max}= 40$ for OCA-WC and $ t_\text{max} = 60$ for SR-WC.
The associated phononic spectral function is computed by
substituting the electronic Green's function by its phononic counterpart. 

In the NRG calculations, the spectral functions are computed through the Lehmann
decomposition, and using the full-density-matrix approach to
approximate the thermal density matrix for temperature $T$. The raw
spectra in form of weighted $\delta$ peaks are broadened using a
log-Gaussian kernel with $\alpha=0.15$ (or $\alpha=0.3$ for mapping out the phase
diagrams) and further with a Gaussian kernel
with a width of order $T$.

\section{Equilibrium}\label{Sec:eq}
\subsection{Phase diagram}\label{Sec:Phase_diag}

\begin{figure}[t]
\includegraphics{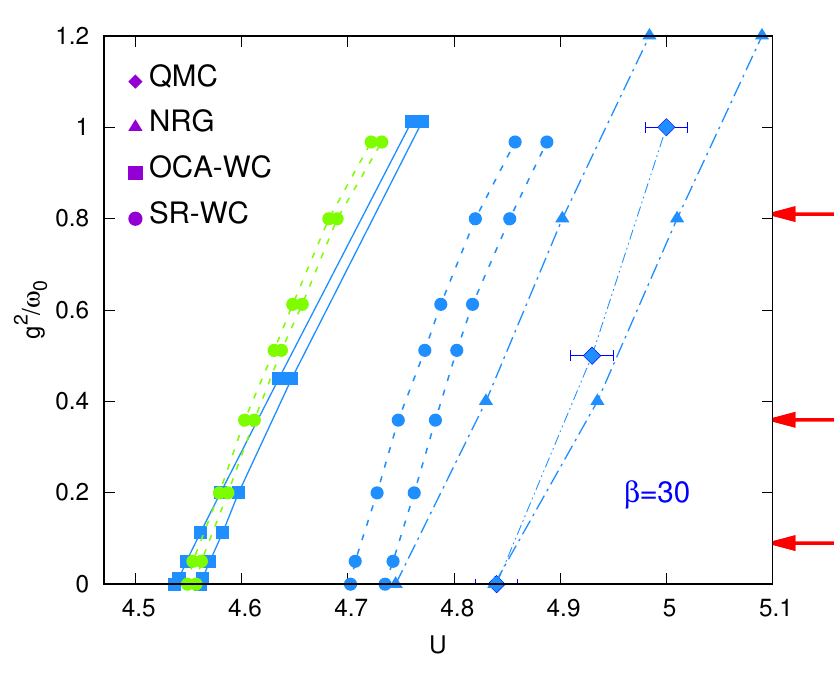}
\caption{DMFT phase diagram of the Hubbard-Holstein model obtained for $\omega_{0}=0.2$ using 
  SR-WC~(blue circles points), OCA-WC~(blue square points), NRG~(blue triangle points)
  and QMC\cite{Werner2007}~(blue diamond points with error bars, $U_{c2}$ only) impurity solvers.
The arrows mark the el-ph couplings for which the analysis
of the spectral functions is presented in Fig.~\ref{Fig:Spectw02}.
The additional green line represents the SR-WC phase boundary at
$\beta=20$. The NRG phase boundary was determined for a larger discretization
 $\Lambda=2.5$ and broadening $\alpha=0.3$ parameter than in the rest of the manuscript.
}
\label{Fig:phasediagram}
\end{figure}

Figure~\ref{Fig:phasediagram} presents the phase diagram of the
Hubbard-Holstein model obtained by different impurity solvers at inverse temperature $\beta=1/T=30$.
The lines delimit the coexistence regime $[U_{\rm c1}, U_{\rm c2}]$ 
for the transition between the correlated metal and the Mott insulator. 
These critical Hubbard interactions depend on the el-ph coupling $g$ and
are renormalized towards larger values upon enhancing $g$. 
This is due to the retarded phonon-mediated el-el attraction
which results in a reduction of the effective Coulomb repulsion.
Intuitively, by increasing the el-ph coupling electrons can excite more phonons
which gives rise to a larger el-el attraction and
a reduction in the repulsive Coulomb interaction.

We note some deviations between the slope of the numerically exact $U_{c2}$-curve from QMC\cite{Werner2007} and the corresponding NRG result. This is a consequence of the NRG truncation at the initial steps of the iteration and the usage of a bigger discretization $\Lambda=2.5$ and broadening $\alpha=0.3$ parameters in the scan of the phase diagram (due to computational cost). While we will use the NRG data as the benchmark in the following discussion, it should be kept in mind that the corresponding spectra involve some approximations in the larger el-ph coupling regime and that these approximations tend to overestimate the metallic character.   

Integrating out the phononic degrees of freedom from the action
obtained from Eq.~\eqref{eq:HamHH} shows that reproducing the spectral
properties of the Hubbard-Holstein model within a purely electronic
system is possible if the effective Hubbard interaction has the
frequency dependence
\begin{equation}
 U_{\rm eff}(\omega) = U - \frac{2 g^{2} \omega_{0}}{\omega_{0}^{2} - {\omega}^{2} }. \label{eq:Ueff}
\end{equation}
In the anti-adiabatic limit $\omega_{0}/W \rightarrow \infty$, this
dynamical Hubbard interaction simplifies to the static value
$U_{\rm eff}=U-2 g^{2}/\omega_{0}$. Away from this limit, however, the
competition between different energy scales leads to
nontrivial low-energy physics. It is thus an interesting problem to
define a static effective Coulomb repulsion which reproduces the
low-energy spectral properties of the original Hubbard-Holstein
model. In Sec.~\ref{sec:effHubbard} we will describe an approach to
calculate this interaction.

The comparison of the phase boundary in Fig.~\ref{Fig:phasediagram}
between SR-WC and OCA-WC reveals that both approaches capture the
renormalization of the metal-insulator transition line.  Already in
the purely electronic model~($g=0$) the coexistence region is
different in both approaches, originating from the different
approximate treatments of charge fluctuations. 
To be precise, these
approximations are: 1) the use of the non-crossing approximation in
the auxiliary space of the slave-rotor method, and 2) the one-crossing
approximation in the OCA-WC formalism. Besides, it is evident that the
coexistence regime of the OCA-WC phase diagram shrinks by enhancing
the el-ph coupling while the coexistence region within the SR-WC
approach is roughly constant as a function of the el-ph coupling. From
now on we will study SR-WC and OCA-WC results at the temperatures
where a decent agreement in the location of the phase-boundary is obtained, namely $\beta=20$
in SR-WC and $\beta=30$ in the OCA-WC.

\subsection{Spectral properties}\label{Sec:Spectral}
In this section, we present a comparison of the equilibrium
spectral functions obtained from the different approximations. The analysis is restricted to weak
and intermediate el-ph couplings, because of the weak~(el-ph) coupling methods,
namely OCA-WC and SR-WCm break down as we approach the bi-polaronic transition.

\begin{figure}[t]
\includegraphics{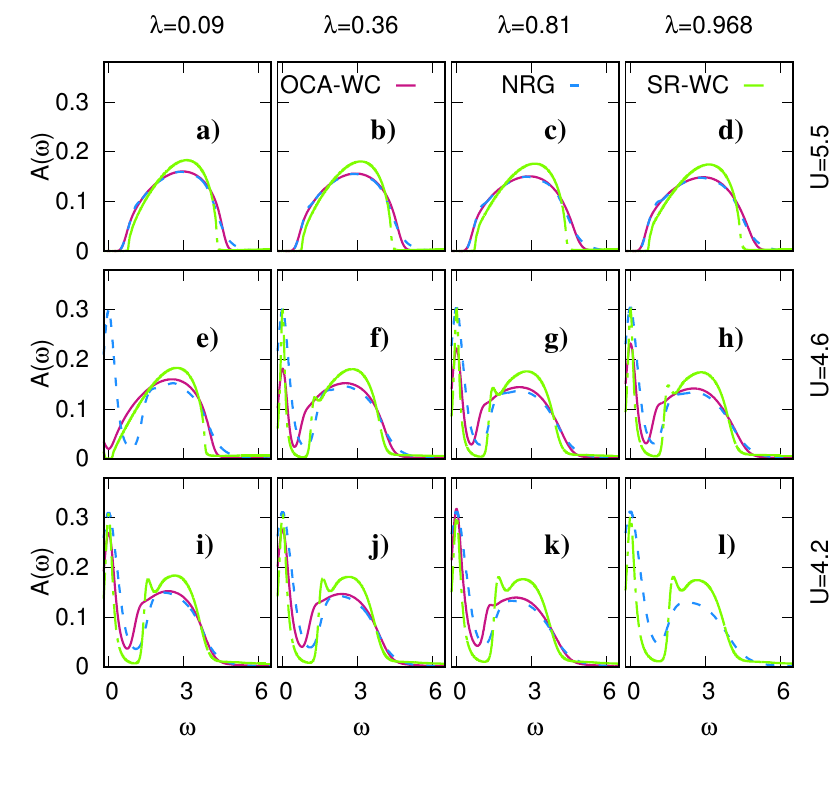}
\caption{ Equilibrium spectral function~$A(\omega)$ obtained from OCA-WC~(solid red lines),
NRG~(dashed blue lines), and SR-WC~(dashed dotted green lines) 
for $U \in \{ 4.2, 4.6, 5.5 \}$,
$g \in \{ 0.134, 0.268, 0.4,0.44 \}$ and fixed phonon frequency $\omega_0=0.2$.
Panels on the same row show results for the same Hubbard interaction,
while the vertically aligned panels correspond to a fixed el-ph coupling.
For $U=4.2$ and $\lambda=0.968$ the OCA-WC calculation fails to converge.
}
\label{Fig:Spectw02}
\end{figure}

\begin{figure}[t]
\includegraphics{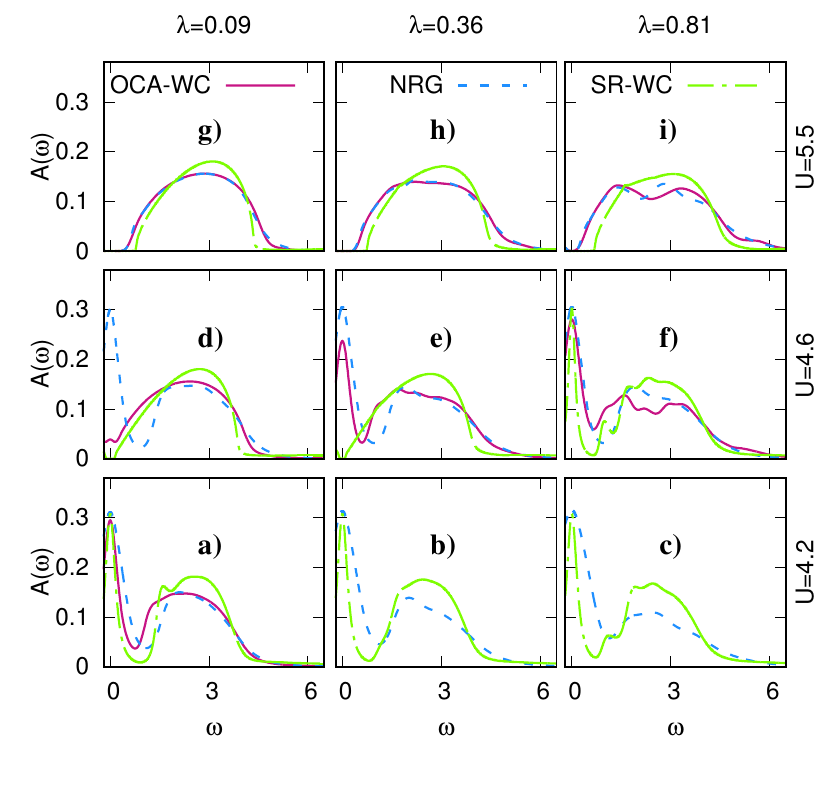}
\caption{ Equilibrium spectral function~$A(\omega)$ obtained from OCA-WC~(solid red lines),
NRG~(dashed blue lines), and SR-WC~(dashed dotted green lines)
for $U \in \{ 4.2, 4.6, 5.5 \}$, $g \in \{ 0.3, 0.6 ,0.9 \}$ and fixed phonon frequency $\omega_{0}=1.0.$
Missing data for the OCA indicate that a converged solution could not be obtained.
}
\label{Fig:Spectw1}
\end{figure}

\textit{Low frequency phonons} The first comparison in Fig.~\ref{Fig:Spectw02}
shows spectra in the Mott insulating and strongly correlated metallic regimes
for increasing el-ph couplings at fixed phonon frequency $\omega_0=0.2$.
In the insulating phase, see the first row in Fig.~\ref{Fig:Spectw02} for $U=5.5$,
due to the strong el-el repulsion, the phonon-mediated deformations of the spectrum are barely noticeable.
The OCA spectral function in this phase
nicely agrees with the reference NRG spectral function for all couplings.
A slight difference can be observed at the edge of the band, where NRG exhibits a slightly broader tail,
which originates from the NRG broadening.
The SR shows a consistent behavior but the bandwidth of the Hubbard band
is smaller due to the symmetry of the employed rotor.\cite{Florens2002}
The comparison of the spectra closer to the metal-to-insulator transition
is complicated due to the fact that the numerical value of the critical
Hubbard interaction $U_c$ differs among the methods, see the second
row in Fig.~\ref{Fig:Spectw02} for $U=4.6$ and the phase diagram in
Fig.~\ref{Fig:phasediagram}. For the weakest el-ph coupling,
$\lambda=0.09$, the NRG calculations exhibit a strongly renormalized
quasi-particle peak, which is not yet manifest in the OCA spectrum,
while the presented SR results are at higher
temperatures, see
Fig.~\ref{Fig:phasediagram}. At the stronger el-ph interactions, $\lambda\geq0.36$, the
quasi-particle peak is present in all approximations, but its weight
is consistently larger in NRG. 

As the el-ph interaction is increased a spectral feature appears at the lower edge
of the Hubbard band and it is most pronounced in the SR-WC, while it
is completely absent in the NRG. The comparison at the lowest
interaction strength $U=4.2$ shows a similar trend, however, the
convergence in the OCA approximation was much slower and we failed to
converge the OCA result for the strongest depicted el-ph interaction
$\lambda=0.968.$

\textit{High frequency phonons} While in the adiabatic limit
($\omega_0/W\ll1$) the vertex corrections are suppressed, as the
phonon energy gets comparable to the electronic energy scale, we
expect that the phonon effects become more pronounced. To demonstrate the
effect on the spectral functions we present a similar comparison as
before, but for
the phonon frequency $\omega_0$ set equal to the hopping $v$, $\omega_0=v=1$,
while keeping the same dimensionless el-ph coupling
$\lambda$, see Fig.~\ref{Fig:Spectw1}. As the el-ph coupling
is increased the deformation of the Hubbard bands becomes more
evident. It leads to a splitting of the Hubbard band into two peaks and we interpret the
lower peak as a polaronic feature. 
This feature is already present in the insulating phase $U=5.5$, where
the agreement between NRG and OCA is reasonably good. However, for the
strongest el-ph interaction $\lambda=0.81$ the
splitting between the peaks substantially differs. Even though one would like to
attribute the higher energy features to additional discrete phonon excitations
the numerical data do not support this picture, since the splitting is
larger than the bare phonon energy $\omega_0$. 
NRG and OCA spectra mainly disagree in the energies of these sidebands.
In contrast, the SR results are different: they show only a shoulder-like feature at the lower
edge of the upper Hubbard band. 

\begin{figure}[t]
\includegraphics{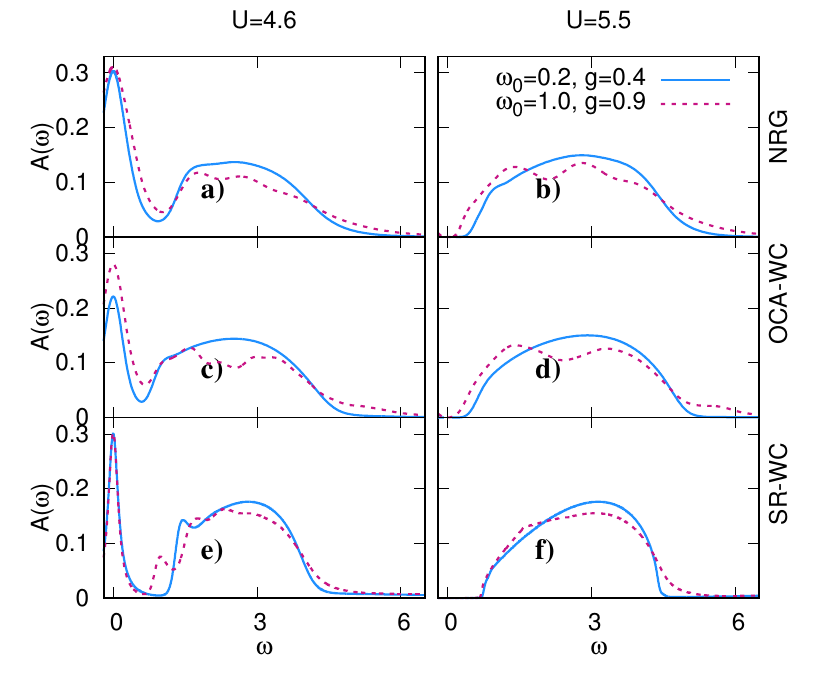}
\caption{Effect of the phonon frequency on the electronic spectral function
for $\omega_{0} \in \{ 0.2,1.0\}$ at $\beta=30$~($\beta=20$ for SR-WC). 
Panels a),c), and e) present spectral densities for $U=4.6$ while panels b), d), and f)
show the corresponding results for $U=5.5$. 
The el-ph coupling strength is fixed at $\lambda=0.81$. }
\label{Fig:Spect2w0}
\end{figure}

The strongly correlated metal at
$U=4.6$ and $U=4.2$ exhibits a rich internal structure of the upper
Hubbard band with several peaks, which become sharper when increasing the
el-ph interaction $\lambda.$ These structures extend the
Hubbard bands to higher energies and therefore systems with the same
dimensionless el-ph coupling $\lambda$ have a larger
bandwidth for larger phonon frequency $\omega_0,$ see also
Fig.~\ref{Fig:Spect2w0}. This is a direct consequence of the fact that
the spectral function is normalized to unity. The different methods do
not agree on the detailed shape of these high-energy features. This
disagreement originates from the different approximations, but also
from the broadening used in the NRG and the finite Fourier window
employed in the calculation of the spectral functions in the OCA
approximations.

In order to illustrate the evolution of the spectral function for different interaction
strengths within a given approximation we present in Appendix~\ref{appendix}
the same data set as in Figs.~\ref{Fig:Spectw02} and \ref{Fig:Spectw1},
but restructured such that each plot shows the evolution of the spectral function
with increasing el-ph interaction for a given approximation. 

\textit{Effect of the phonon frequency} To demonstrate the effect of
the phonon frequency $\omega_0$ on the electronic properties we
compare the spectral function $A(\omega)$ for $\omega_0=0.2$ and
$\omega_0=1.0$ within different approximations at a fixed
dimensionless el-ph coupling $\lambda$ in
Fig.~\ref{Fig:Spect2w0}. For the strongly correlated metal the main
effect of the increased phonon frequency $\omega_0$ is the enhancement
of polaronic effects leading to the internal structures in the
Hubbard band and the associated increase in the bandwidth. In
NRG and OCA the quasi-particle weight is increased for high-frequency
phononic modes, while in the SR it remains almost constant. In the
Mott insulating phase, namely $U=5.5$, the NRG and OCA shows a
renormalization of the Hubbard gap, while in the SR this effect
is much smaller. 

\subsection{Renormalized phonon frequency}

\begin{figure}[t]
\includegraphics{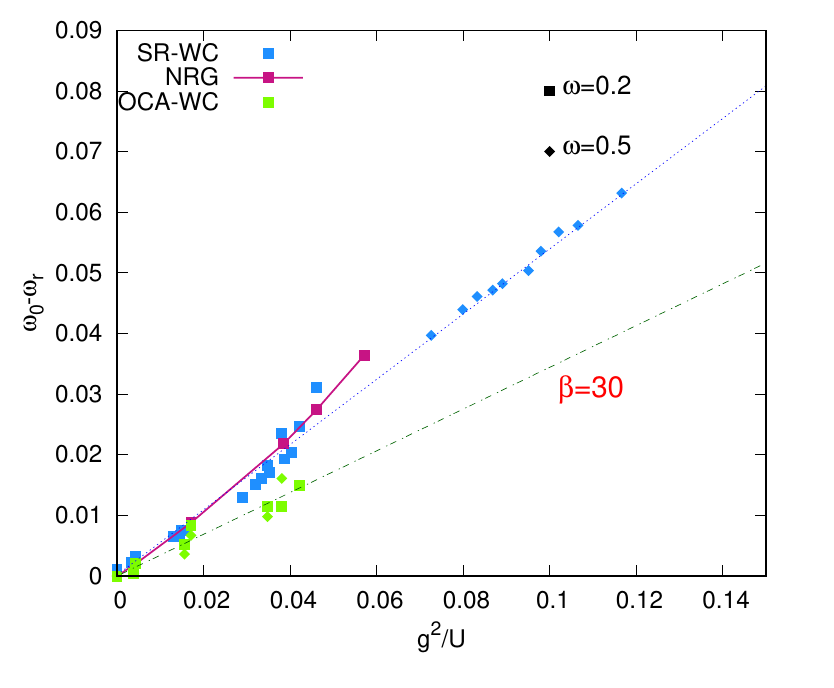}
\caption{Renormalized phonon frequency obtained from various approximations. 
The dashed~(dot-dashed) blue~(dark green) line is the linear fit of the SR-WC~(OCA-WC)
results for $\omega_{0}=\{ 0.2, 0.5 \} $ and $U<U_{\rm c}$.
SR-WC results are computed at $\beta=20$.}
\label{Fig:Renormphonon}
\end{figure}

As a result of the feedback between the electronic and phononic
subsystems, the effective phonon frequency~($\omega_{\rm r}$) is
renormalized. In this section, we compare how well the
renormalization of the phonon frequency is captured within each of the
approximations. The renormalized frequency is extracted from the
position of the peak in the phonon spectrum.
We start with the observation that within SR-WC one finds an explicit scaling for the phonon softening, 
\begin{equation}
 \omega_{0} -\omega_{r}= \alpha \frac{g^{2}}{U},\label{eq:wr}
\end{equation}
which originates from the interaction between charge fluctuations (as described by the rotor) and phonons.
Here, $\alpha$ is a proportionality factor which depends on the model parameters.
The detailed derivation of this scaling is presented in Appendix~\ref{Supp:renormphonon}.
The basic assumption is that charge fluctuations are reduced, as expected within the Mott phase,
which leads to the emerging small parameter $2g^2/(U\omega_0),$ see also Appendix~\ref{Supp:renormphonon}.
In the zeroth order of the charge-phonon coupling, we would obtain $\alpha=2$,
and therefore the value of this fitting parameter can be taken as a measure for the effective interaction between the charge and phonon sectors. 
In Fig.~\ref{Fig:Renormphonon}, we illustrate to which extent Eq.~(\ref{eq:wr})
holds within the SR-WC, OCA-WC, and NRG approximations. 
A roughly linear dependence between $\omega_{0} -\omega_{r}$ and $\frac{g^{2}}{U}$ is found in all methods,
although the associated slopes for NRG and SR-WC are larger~($\alpha_{\rm SR/NRG} \approx 0.5$)
than for the OCA-WC formalism~($\alpha_{\rm OCA} \approx 0.3$). 
This difference can be attributed to the strong-coupling diagrammatic nature of the OCA,
which underestimates the charge fluctuations responsible for the phonon softening.
The deviation from the linear fitting for SR-WC and NRG is more evident when
$g^{2}/U $ is comparable or larger than $\omega_{0}$.
This behavior is rooted in the moderate interplay between the local
charge-fluctuations and the phonon displacement.

\subsection{Quasi-particle weight and effective static interaction}\label{sec:effHubbard}
The effective low-energy physics in the strongly correlated metal is
determined by the quasi-particle weight
$Z=[1-\partial \Sigma/\partial \omega|_{\omega=0}]^{-1},$ which in the DMFT context
is also the inverse of the effective mass $Z=m/m^*.$ The effect of the
el-ph interaction on the quasi-particle weight $Z$ is
twofold: a) the phonon-mediated effective interaction is screened, see
Eq.~(\ref{eq:Ueff}), and the reduced static interaction leads to an 
enhanced quasi-particle weight $Z$, b) the dressing of the
quasi-particle with the phonon cloud leads to an enhanced effective
mass $m^*$ or equivalently to a reduced quasi-particle weight $Z.$
In the atomic limit, the renormalization is given by the Lang-Firsov
factor $Z_B=\exp \big(-g^2/\omega_0\big)$.\cite{mahan2000} The overall effect of
the el-ph interaction on the low-energy physics is a
non-trivial problem resulting from the competition between these two
mechanisms. Here we will follow Ref.~\onlinecite{Sangiovanni2005},
where it was proposed that the low-energy physics of the
Hubbard-Holstein problem can be effectively described by a purely
electronic system with a renormalized interaction and that retardation
effects only affect the high-energy region of the spectrum.

Due to the finite propagation time $t_\text{max}$, the evaluation of the quasi-particle weight
from the derivative of the self-energy becomes a tedious task, and the subsequent
non-equilibrium analysis exacerbates this problem. Here we instead propose an analysis based
on the integral over the low-energy part of the photo-emission spectrum (PES) $I(\omega)$.
The latter is computed as \cite{freericks09}
\begin{equation}
I(\omega)= \text{Im} \int \frac{{\rm d}t_{1} {\rm d}t_{2}}{2 \pi} S(t_{1}) S(t_{2})
e^{\mathrm{i} \omega(t_{1}-t_{2})} G^{<}(t_{1}-t_{2}),
\end{equation}
for a Gaussian probe pulse with the time resolution $\delta$ given by
$S(t) =\exp(t^{2}/\delta^{2})$, where $\delta$ is set to be smaller
than the phonon period~($2\pi/\omega_{0}$) and we have used the time-translational invariance. In order to have a practical
measurement of the quasi-particle weight also out of equilibrium we
use the low-energy integral $I=\int_{-0.2}^{0.2} I(\omega) d\omega$ as the matching condition
between the el-ph coupled system and the effective electronic
system. In other words, the effective interaction of the purely electronic Hubbard
model is determined by matching the low-energy integral $I$ to the result
obtained from the Hubbard-Holstein model.
 
\begin{figure}[t]
\includegraphics{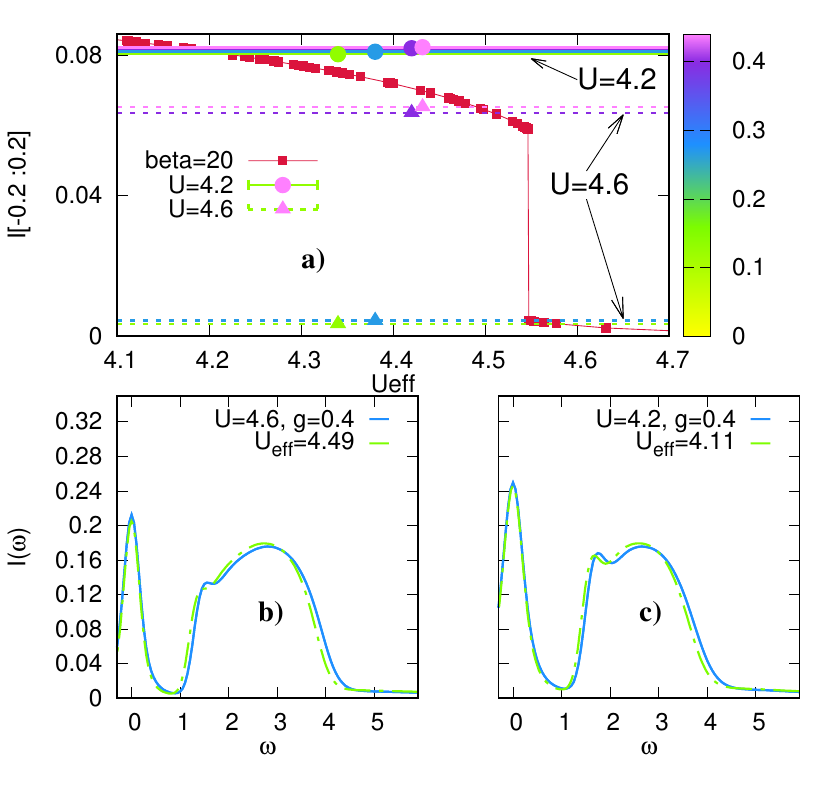}
\caption{ a) SR-WC results for the integral over the low-energy PES
$I=\int_{-0.2}^{0.2} I(\omega) d\omega$ obtained from the purely electronic model
(red line) and the electron-boson coupled system (horizontal lines) for $U \in \{4.2,4.6\}$
and different el-ph couplings $g \in \{0.134,0.268,0.4,0.44\}$
, whose values are given in the color-bar. 
Comparison of the spectral function for the Hubbard-Holstein model (blue full line)
at $g=0.4$ and $U=4.6$~(b) and $U=4.2$~(c) and the Hubbard model
with the effective interaction $U_\text{eff}\approx 4.49$~(b) and $U_\text{eff}=4.11$~(c).
The phonon frequency is $\omega_0=0.2$.
}
\label{Fig:effHubbardSR}
\end{figure}

\begin{figure}[t]
\includegraphics{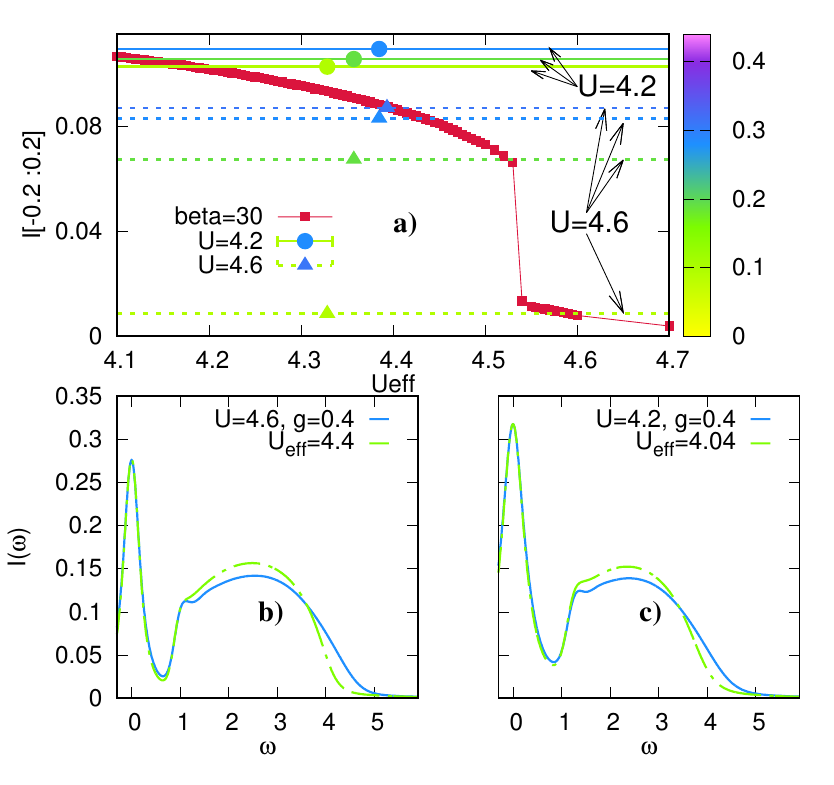}
\caption{ Similar analysis as in Fig.~\ref{Fig:effHubbardSR} for the
  OCA-WC method. The effective interaction for $U=4.6, g=0.4$ is
  $U_\text{eff}\approx 4.4$~(b), while for $U=4.2,g=0.4$ it is given by
  $U_\text{eff}=4.04$~(c). The phonon frequency is $\omega_0=0.2$. 
  }
\label{Fig:effHubbardOCA}
\end{figure}

In Fig.~\ref{Fig:effHubbardSR} and Fig.~\ref{Fig:effHubbardOCA} we
present the analysis for SR-WC and OCA-WC, respectively. The
interaction strengths in the Hubbard-Holstein case are $U=4.6$,
corresponding to the Mott insulating phase without el-ph
coupling, and $U=4.2$, which is a strongly correlated metal without
el-ph coupling. The increase of the el-ph coupling
$g$ leads to an enhanced integral  $I$ over the quasi-particle, see
Fig.~\ref{Fig:effHubbardSR}(a) and
Fig.~\ref{Fig:effHubbardOCA}(a). The effective electronic interaction
$U_\text{eff}(g,U)$ is then determined by matching the low-energy integral
$I$ from the Hubbard-Holstein problem with the one obtained from the
Hubbard model $I(U_\text{eff})=I(g,U).$ As can be seen from
Fig.~\ref{Fig:effHubbardOCA}(a) this condition is not always
fulfilled since for the Mott state the low-energy integral $I$ from
the Hubbard-Holstein model can lie within the jump induced by the
first order MIT. A direct comparison of the PES is presented in
subplots b), c) of Fig.~\ref{Fig:effHubbardSR} and Fig.~\ref{Fig:effHubbardOCA} for the
SR-WC and OCA-WC methods, respectively. These panels confirm the main result of
Ref.~\onlinecite{Sangiovanni2005} that the low-energy spectrum of the
two models is practically identical. 
This serves as a confirmation
that the integral over the quasi-particle peak $I$ is a reliable matching
condition for the low-energy physics of the Hubbard-Holstein and
Hubbard model. In the following section, we will use this insight for 
an analysis of the non-equilibrium
dynamics to see how the low-energy physics can be modulated by an
external perturbation and to check if one can always find a purely electronic system
that matches the low-energy physics of the Hubbard-Holstein problem.

\section{Non-equilibrium }\label{sec:neq}

We now turn to the study of non-thermal IMTs, by investigating the temporal response of the system
after a sudden quench of the el-ph coupling in the adiabatic
limit ($\omega_{0}=0.2$). We abruptly increase the coupling parameter from $g=0$ to a nonzero final value. Using
this protocol, we investigate the transient properties of systems 
close to the metal-to-insulator transition and in the coexistence regime.

\subsection{Double occupation and kinetic energy}\label{subsec:Ekind}

We will first consider the time evolution of the double occupancy and kinetic energy for initial states in the correlated metallic~($U=4.4, g=0$) or
insulating~($U=4.6, g=0$) phase. 
After switching on the el-ph coupling, the phonons screen the Hubbard
repulsion, see Eq.~(\ref{eq:Ueff}), and consequently, the effective
repulsion is reduced. 
The nontrivial question is whether the system relaxes into a new thermal state and whether the screening can induce an IMT.

\subsubsection{Correlated metal - $U=4.4$} 
\begin{figure}[t]
\includegraphics{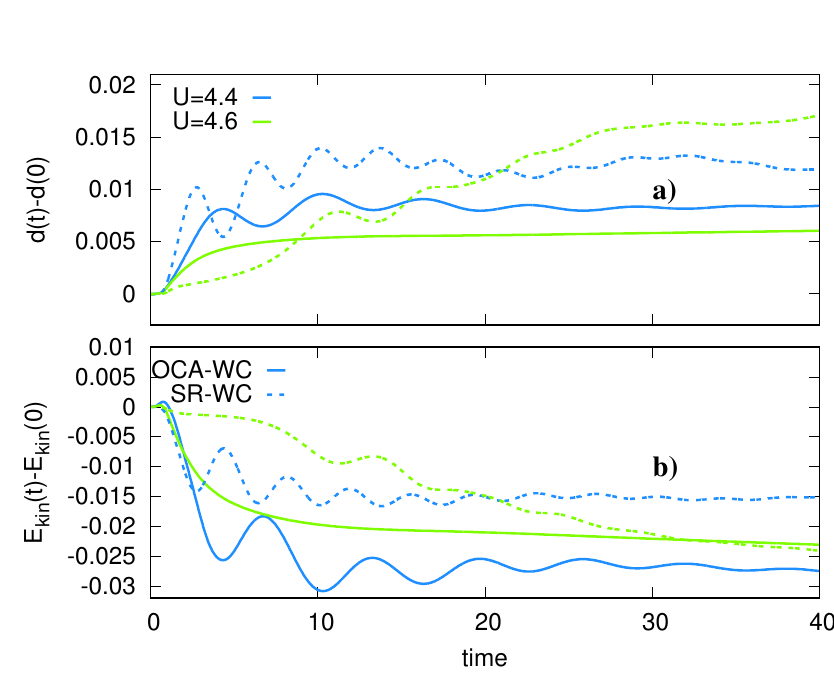}
\caption{Time-dependent double-occupancy~a) and kinetic energy~b)
within OCA-WC~(solid line) and SR-WC~(dashed line) at
$U \in \{4.4, 4.6\}$, $\omega_{0}=0.2$, $g=0.44$ and $\beta=\{ 20$~(SR-WC), $30$~(OCA-WC)$\}$. 
}
\label{Fig:NEekin}
\end{figure}

The dynamics after a quench to
$g=0.44$ is shown for OCA-WC and SR-WC in Fig.~\ref{Fig:NEekin}.  As
expected from the reduction of the effective interaction, the double
occupation increases, and the kinetic energy is suppressed as the
system evolves towards a putative metastable state. The transient
evolution is characterized by strong oscillations which can be linked
to the following two processes: 

a) Creation of holon-doublon pairs by the quench. In particular, for
the initial metallic state, the oscillation frequency is determined by
the energy difference between the quasi-particle band and the Hubbard
bands, and thus the oscillations can be associated with excitations
between these bands.  This picture is also confirmed by the occupation
dynamics, which exhibits long-lived oscillations on these two energy
scales, namely from the lower to the upper Hubbard band and from the
quasi-particle peak to the upper Hubbard band. This scenario is further 
supported by the fact that the spectral function is almost fixed for
$t>15$. 

b) The creation of holon-doublon pairs leads to enhanced fluctuations of
the phononic field and increases the polaronic tendencies of the
system.

\begin{figure}[t]
\includegraphics{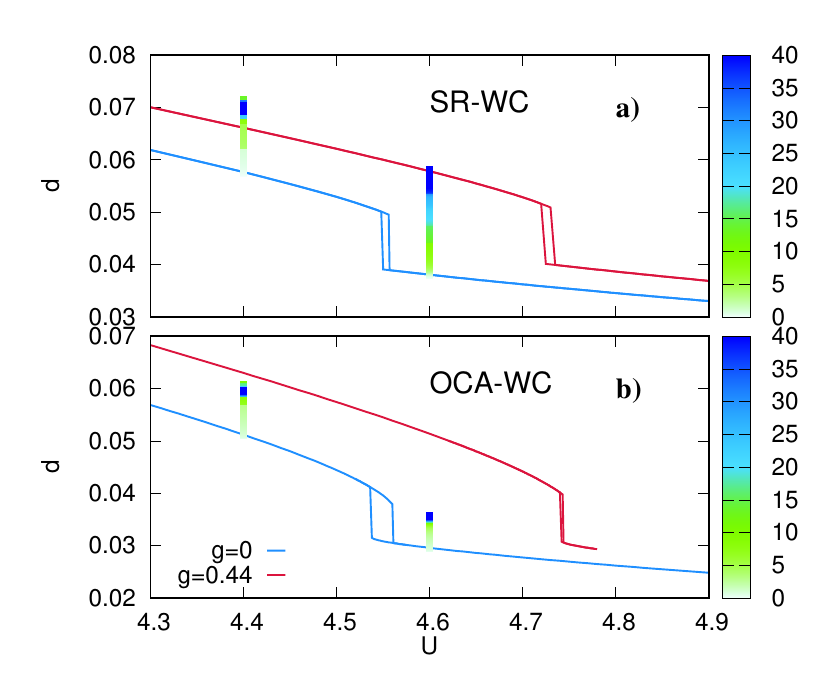}
\caption{Time evolution of the double-occupancy~(colored bar)
  within SR-WC~a) and OCA-WC~b) at $\beta=\{ 20$~(SR-WC),
  $30$~(OCA-WC)$\}$. The color box denotes time. Blue lines exhibit
  the phase diagram of the Hubbard model and red lines present the
  associated phase diagram of the Hubbard-Holstein model at
  $\omega_{0}=0.2$ and $g=0.44$. 
  }
\label{Fig:docc_on_phase}
\end{figure}

\begin{figure}[t]
\includegraphics{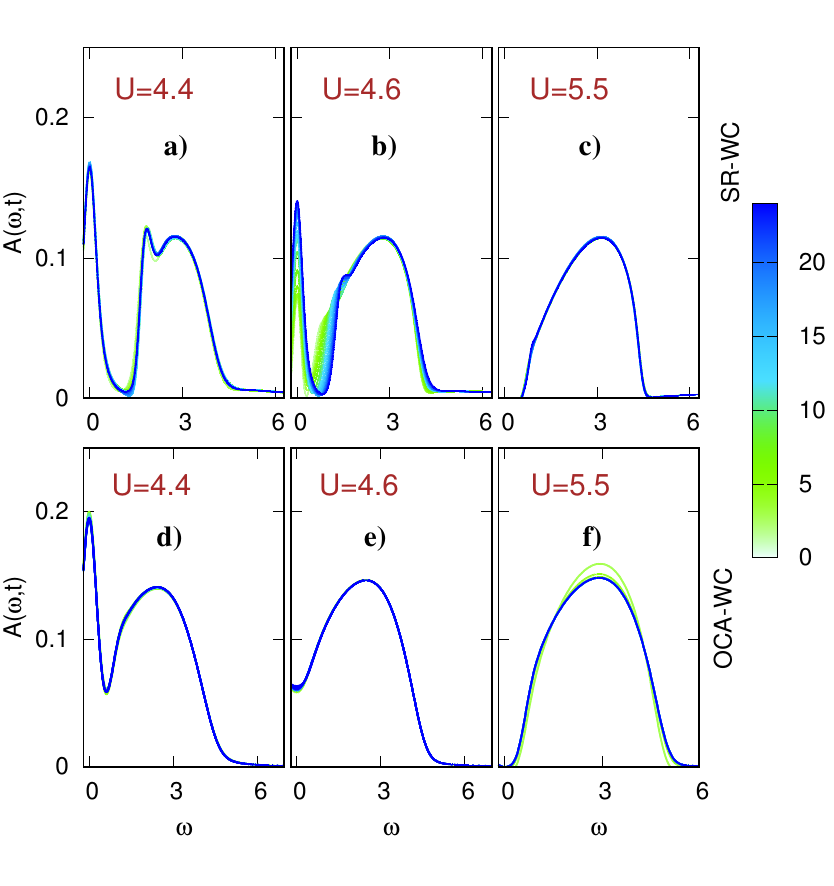}
\caption{a-c) SR-WC results for the time-dependent spectral functions~($A(\omega,t)$)
at $U \in \{ 4.4, 4.6, 5.5 \}$, $\omega_{0}=0.2$ and final el-ph coupling $g=0.44$.
d-f) Analogous OCA-WC results as
a function of time at $U \in \{ 4.4, 4.6, 5.5 \}$, $\omega_{0}=0.2$
and final el-ph coupling $g=0.44$. The color box denotes time.
}
\label{Fig:Spect_NE}
\end{figure}

In order to compare the non-thermal state after the quench to the
associated equilibrium states we present the time evolution of the
double occupancy as a function of time~(bars) together with a plot of the
equilibrium hysteresis region of the double occupancy in the Hubbard~(initial
Hamiltonian, solid blue line) and Hubbard-Holstein model~(final
Hamiltonian, red solid line), see Fig.~\ref{Fig:docc_on_phase}. The
latter correspond to equilibrium results at $\beta=20$ (SR-WC) and
$30$ (OCA-WC) and the 
final $g$. In both SR-WC and OCA-WC, the initial trend is an increase
of the double occupancy and an approach to the equilibrium value of
the final el-ph coupled Hamiltonian after the quench. In the initial metallic phase~($U=4.4$),
the two methods also agree for longer times, where the double occupancy is slightly
enhanced and the quasiparticle peak in
the spectral function is reduced, see Fig.~\ref{Fig:Spect_NE}. However in the
long time limit the SR exceeds the equilibrium value in contrast to the OCA results.
For the initial insulating phase at longer
times these two methods start to quantitatively deviate: SR-WC
shows a stronger increase in the double occupancy than OCA-WC and the
transient value even exceeds the equilibrium reference.
Note that the solid red line in Fig.~\ref{Fig:docc_on_phase} is the
reference system at $\beta=20$ or $30$ and not the expected final
thermal state of the system. While this implies that the
associated thermal states have higher effective temperatures, further validation of this scenario requires
longer simulation times. 
The difference in the double occupancy for
long times is not so surprising since the time evolution is governed
by a subtle interplay of various factors, like the reduction of the
effective el-el interaction, the increase in the charge
fluctuations and the renormalization of the phonon frequency. The two
approximations represent a different competition between these effects
and therefore it is hard to give a quantitative description of the
expected final thermal state and the effective temperature.
Nevertheless, the qualitative behavior is consistent: the non-adiabatic
switching of the el-ph coupling reduces the effective interaction  of
the system which thus relaxes into a more metallic state. In
Sec.~\ref{sec:NEspect} we will show that the quasi-equilibration of
our transient state survives at low energies
and we consequently will associate an effective temporal temperature to this energy range.
Our results will show that the reduction of the static el-el repulsion in systems which are initially in the correlated metal phase is not dramatic, see Sec.~\ref{sec:NEeffectiveU}. Therefore in strongly correlated
metals, we identify the largest contribution as coming from the charge
fluctuations and the renormalized phonon frequency.

\subsubsection{Mott insulator}
If we start in the insulating phase of the el-ph uncoupled
system~($U=4.6$), the transient evolution exhibits an increase in the
double occupancies, both within the OCA-WC and SR-WC description, see
Fig.~\ref{Fig:NEekin} a) and Fig.~\ref{Fig:docc_on_phase}. In the
OCA-WC simulation this enhancement is gradual and monotonic, which can
be explained by the small reduction of the Hubbard interaction as a
result of the coupling to phonons, see Sec.~\ref{sec:NEeffectiveU}. In
contrast, for SR-WC the increase of the double occupancy is
accompanied by shallow oscillations which are a
consequence of two processes: a) the build-up of the coherent
quasiparticle peak, and b) a pronounced renormalization of the phonon
frequency due to the appearance of conducting electrons. The
double occupation increases almost to the reference value of the
Hubbard-Holstein model with $g=0.44$, see Fig.~\ref{Fig:docc_on_phase}~a), and
indicates that the evolution of the system is towards the correlated
metallic phase. In the absence of el-el
interactions~\cite{Murakami2015} a roughly similar timescale governs
the coherent oscillations of local observables which thermalize in
less than ten cycles. In the Hubbard-Holstein system, however,
investigating the full
thermalization is numerically demanding, and will not be attempted here. 

\subsection{Time-dependent spectral function}\label{sec:NEspect}
As we have discussed in Sec.~\ref{sec:effHubbard}, in equilibrium, the height of the
quasiparticle peak can be reproduced by a purely electronic
system by introducing a properly renormalized interaction. We now 
apply an equivalent protocol also out of equilibrium to investigate 
the time dependence of the effective interaction. Figure~\ref{Fig:Spect_NE} plots 
the time-dependent spectral functions
at various Hubbard interactions for the SR-WC and OCA-WC approximations.

\paragraph{Initial metallic phase} Figures~\ref{Fig:Spect_NE} a) and
d) present the temporal evolution of the spectral functions for
$U=4.4$ and final el-ph interaction~$g=0.4$. The initial spectrum has
a three-peak structure and the relative weight of the quasi-particle
band and the Hubbard bands change weakly after the quench. This
redistribution of spectral weight is accompanied by an enhancement of
the quasiparticle peak as the static Coulomb
repulsion is effectively reduced. At $t \gtrsim 15$ 
the phonon cloud dresses the formed polarons, and subsequently, the
height of the quasiparticle peak is decreasing. This is also
accompanied by slow oscillations of the double occupancy as seen in
Figs.~\ref{Fig:NEekin} a) and \ref{Fig:docc_on_phase}. 
Whether the polaron dressing effect dominates the
reduced interaction at longer times is an interesting question which we leave
to future investigations.

\begin{figure}[t]
\includegraphics{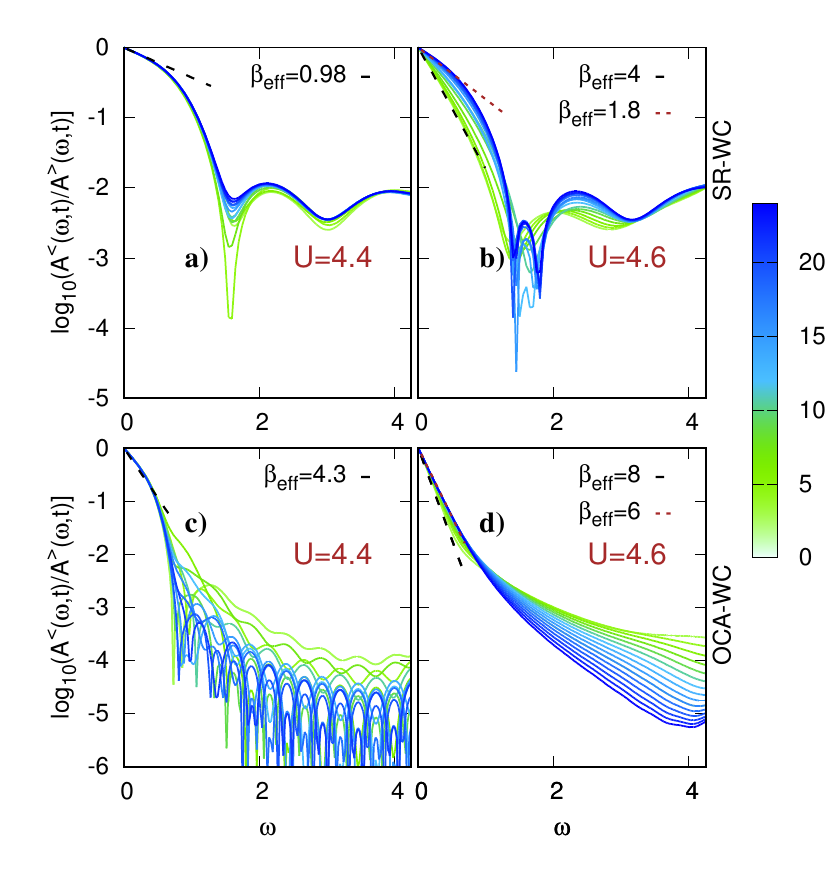}
\caption{a-b) SR-WC results for $A^{<}(\omega,t)/A^{>}(\omega,t)$
at $U \in \{ 4.4, 4.6\}$, $\omega_{0}=0.2$ and final el-ph coupling $g=0.44$.
c-d) Analogous OCA-WC results as
a function of time at $U \in \{ 4.4, 4.6 \}$, $\omega_{0}=0.2$
and final el-ph coupling $g=0.44$.
The color box denotes time.
Dashed black and brown lines are low-energy linear fits of $\exp(-\beta_{\mathrm{eff}}\omega)$ to $A^{<}(\omega,t)/A^{>}(\omega,t)$
at $t=1.5$ and $t=24$, respectively.
}
\label{Fig:NE_fluc}
\end{figure}

The non-thermal nature of the transient state is further evidenced
through the ratio between the non-equilibrium spectral functions of
the occupied~($A^{<}$) and unoccupied~($A^{>}$) states as shown in
Fig.~\ref{Fig:NE_fluc}.  For a thermal state the
fluctuation-dissipation theorem~\cite{Weber1956} requires that
\begin{equation}\label{eq:flucdiss}
 \frac{A^{<}(\omega,t)}{A^{>}(\omega,t)}= e^{-\beta_{\mathrm{eff}} \omega},
\end{equation}
where $T_{\rm eff}=1/\beta_{\mathrm{eff}}$ is the effective temperature of the equilibrated system. 
In Fig.~\ref{Fig:NE_fluc} a) and c) we plot this ratio at
$U=4.4$ for SR-WC and OCA-WC, respectively. At low energies, both the SR-WC and OCA-WC exhibit an almost negligible
transient response and the linear fits to Eq.~(\ref{eq:flucdiss})
yield $\beta_{\mathrm{eff}}=0.98$~(SR-WC) and $\beta_{\mathrm{eff}}=4.3$~(OCA-WC). 
Both approximations consistently have substantially higher low-energy effective temperatures 
than in the initial state, where $\beta=20$~(SR-WC) and $\beta=30$~(OCA-WC). 
Figure~\ref{Fig:NE_fluc} furthermore shows that at $\omega\gtrsim 0.8$ the energy distribution
function is non-thermal. 
In the $U=4.6$ case, OCA-WC exhibits a time-dependent distribution which is consistent with cooling of doublons in the energy region of the Hubbard bands. 
In contrast, the SR-WC distribution changes mainly in the quasi-particle region and around the edges of the Hubbard bands, and shows a rather robust partial inversion of the population in the Hubbard band region.

In the following, we will classify the initial insulating states of the
Hubbard model into two categories which are distinguished by whether
or not their el-el repulsion is lager~(smaller) than the critical
interaction of the thermal electron-phonon coupled
system~($U^{\rm HH}_{\rm c2}\approx 4.72$ for SR-WC and
$U^{\rm HH}_{\rm c2}\approx 4.74$ for OCA).

\paragraph{Initial deep Mott insulating phase }
As a representative of the first category~($U>U^{\rm HH}_{\rm c2}$) we show the
time-dependent spectral function at $U=5.5$ in
Figs.~\ref{Fig:Spect_NE} c) and f) for the OCA-WC and SR-WC,
respectively. It is evident that due to the small charge fluctuations
in this Mott insulating phase, the transient modulation of the
el-ph coupling can hardly mediate low-energy excitations. On
energy scales of the order of $U$, OCA-WC yields a redistribution of the band
which is quickly damped. This response is not significant in SR-WC,
being barely noticeable in Fig.~\ref{Fig:Spect_NE} c). This can be
partially traced back to the shortcomings of the SR-WC in capturing the correct Hubbard bands.

\begin{figure}[t]
\includegraphics{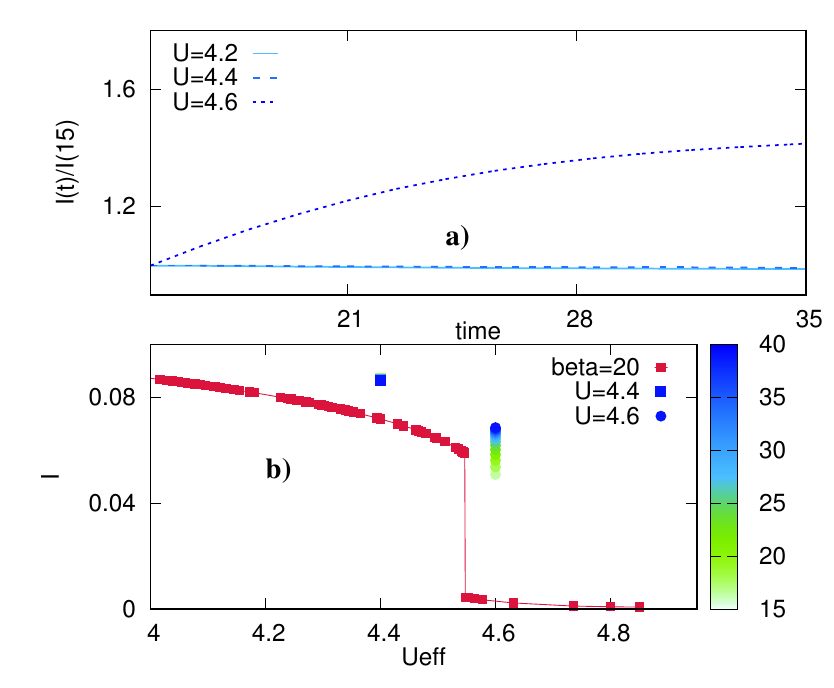}
\caption{ 
a) SR-WC results for the time-dependent integral over the low-energy PES
$I(t)=\int_{-0.2}^{0.2} I(\omega,t) d\omega$ over the associated value at $t=15$
as a function of time at $U \in \{ 4.2, 4.4, 4.6 \}$. 
b) SR-WC results for the integral over the low-energy PES obtained
from the purely electronic model (red line) and the electron-boson coupled system (colored bars)
for $U \in \{4.4,4.6\}$ , $\omega_{0}=0.2$, $g=0.44$ and $\beta=20$.
}
\label{Fig:NEeffHubSR}
\end{figure}

\paragraph{Initial Mott insulating phase close to MIT} 
The closer the Hubbard interaction is to $U^{\rm HH}_{\rm c2}$ the
more the low-energy density varies. For ($U<U^{\rm HH}_{\rm c2}$) we present results at
$U=4.6$ in Figs.~\ref{Fig:Spect_NE} b), and e). In this
parameter regime, the two-peak insulating spectrum of the Mott insulator
gives way to the formation of a quasiparticle peak which grows more
dramatically in SR-WC than in OCA-WC. In this regime, the band
renormalization is recognizable in both SR-WC and OCA-WC. While in the
former approximation the major redistribution of the spectral density
occurs at the band edges, in the latter approach the middle of the Hubbard
band exhibits the strongest renormalizations. The renormalization of
the spectral function in SR-WC facilitates the build-up of the
quasiparticle peak as even small el-ph excitations can
assist the process. In OCA-WC the transferred energy to
accumulate low-energy spectral densities should be of order $W/2$ since we have to excite holon/doublon pairs to enhance the phonon fluctuations.
In the adiabatic regime, this amount of energy is mainly accessible
through multi-phonon processes with a low probability of excitations as
the charge-fluctuations in OCA-WC are suppressed, see also
discussions in Sec.~\ref{Sec:eq}. 

The ratio between occupied and unoccupied states provides information about the non-thermal pathway of the IMTs at $U=4.6$, see Fig.~\ref{Fig:NE_fluc}~b) and d).  Both
the OCA-WC and the SR-WC results exhibit exponential behaviors at low-energies
and their associated effective temperatures vary in time.  This
change of the effective temperature is stronger 
in SR-WC than in 
OCA-WC as the growth of the quasi-particle height is more pronounced in
the former approximation, see also Fig.~\ref{Fig:Spect_NE}.  
At higher-energies, 
SR-WC presents drastic changes around
the band-edge reflecting the strong redistribution of the spectral
weight to the quasi-particle peak, while OCA-WC exhibits considerable
changes at $W/2$ resembling the cooling and doublon/holon recombination dynamics 
expected in a metallic system. 

\subsection{Quasi-particle weight and effective static interaction}\label{sec:NEeffectiveU}

To further investigate the low-energy excitations of the system, we
employ the matching condition introduced in Sec.~\ref{sec:effHubbard} to
analyze the transient effective el-el interaction. The generalization of the PES to the nonequilibrium condition is given by 
\begin{equation}
I(\omega,t)= \text{Im} \int \frac{{\rm d}t_{1} {\rm d}t_{2}}{2 \pi} S(t_{1}) S(t_{2})
e^{\mathrm{i} \omega(t_{1}-t_{2})} G^{<}(t+t_{1}, t+t_{2}).
\end{equation}

\begin{figure}[t]
\includegraphics{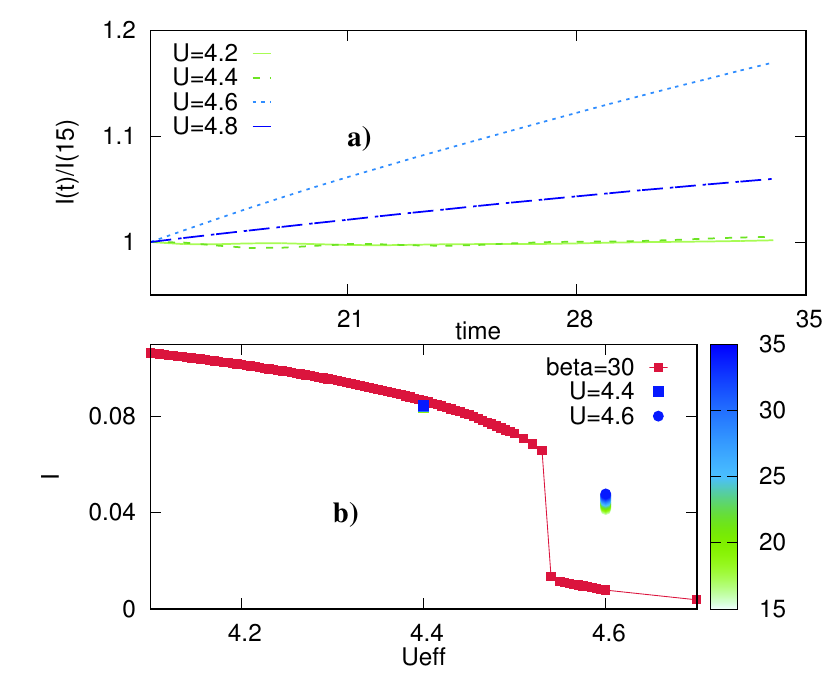}
\caption{a) OCA-WC results for the time-dependent integral over
the low-energy
PES $I(t)=\int_{-0.2}^{0.2} I(\omega,t) d\omega$ over the associated value at $t=15$
as a function of time at $U \in \{ 4.2, 4.4, 4.6, 4.8 \}$, $\omega_{0}=0.2$, $g=0.44$
and $\beta=30$. b) OCA-WC results for the integral over the low-energy PES obtained
from the purely electronic model (red line) and the electron-boson coupled system (colored bars)
for $U \in \{4.4,4.6\}$ and $\beta=30$.}
\label{Fig:NEeffHubOCA}
\end{figure}

The lower panels of Figs.~\ref{Fig:NEeffHubSR},\ref{Fig:NEeffHubOCA} present the
time-dependent integral over the low-energy PES
$I(t)=\int_{-0.2}^{0.2} I(\omega,t)d\omega$ as a function of
time~(bars). In the strongly correlated metal~($U=4.4$) the change in
the quasiparticle weight is small after $t>15$ and the
matching condition suggests a very slow evolution of the effective
Hubbard interaction towards the thermal value.
However the non-thermal
trajectories are very distinct especially for short times in
both approximations. Within OCA-WC the quasi-particle weight 
increases and therefore the effective Hubbard interaction is reduced
as a function of time, which demonstrate the dominant role of screening
of the Coulomb repulsion due to the formation of the phonon cloud.
Within SR-WC the effective el-el repulsion is
reduced considerably at short times and the later time evolution
exhibits a small but gradual decrease of the quasiparticle weight,
which can be related to the dressing of the quasi-particles.

In the Mott-insulating phase, $U<U^{\rm HH}_{c2}$, at $U=4.6$ the
picture is quite different for SR-WC and OCA-WC. In the former, the quasi-particle weight is strongly modified suggesting a strong
reduction of the effective el-el interaction as a function
of time. In this regime, the system traverses the first-order phase transition on a non-thermal path.
In this regime, the phonon screening is the dominant process leading to a strong redistribution of spectral weight to a metal-like PES. 
The OCA-WC shows a similar trend, but without a complete
switching from insulator to metal and in fact a rather small 
increase in the quasi-particle weight. We have checked that within the
OCA-WC approximation on the reachable timescales the full transition
cannot be achieved no matter how close the initial state is to the
critical interaction. The overestimation of the insulating nature of
the state is a well-known artifact of the NCA and OCA approximation in
equilibrium and the above behavior might be a non-equilibrium
manifestation of this artifact.

In addition, comparing the ratio $I(t)/I(15)$ of both SR-WC and
OCA-WC, see upper panels of
Figs.~\ref{Fig:NEeffHubSR},\ref{Fig:NEeffHubOCA}, also suggests that the
increase of the quasi-particle weight is more pronounced for 
Hubbard interactions $U^{\rm H}_{c2} < U <U^{\rm HH}_{c2}$,
where $U^{\rm H(HH)}_{c2}$ is the critical interaction in the
Hubbard~(Hubbard-Holstein) model. We thus conclude that
transitions from Mott insulators to non-thermal correlated metals are
achievable in this range of interactions. Note however that the
relative change is significantly larger in the SR-WC approach than in
the OCA-WC (different $y$-axis scale on both plots).

\section{Conclusions}
\label{sec:Con}

We have employed the DMFT framework to investigate electron-phonon
coupled systems described by the Hubbard-Holstein model in the weak el-ph coupling regime, both in and out of
equilibrium. One purpose of this work was the comparison between different impurity solvers: 
OCA-WC, SR-WC, and NRG. In equilibrium, we have used the NRG results as reference data to assess the validity
of the OCA-WC and SR-WC solvers in various parameter regimes and to reveal
the equilibrium properties of the spectral
function. 

We have found that in the Mott insulating regime the spectral properties obtained within OCA-WC 
are in excellent agreement with the NRG
counterparts, while SR-WC exhibits shortcomings in capturing the
shape of the Hubbard bands. We have pointed out that this drawback is a
consequence of employing the non-crossing approximation in the
auxiliary Hilbert space of the problem. Nevertheless, the presented
SR-WC diagram features a more accurate metal-Mott insulator phase boundaries
than the result obtained with OCA-WC. Here, we have to note that the fudge parameter is determined to have the correct MIT at $g=0$ and the remaining phase boundary is calculated without further adjustment. The underestimation of the
critical Hubbard interactions in the OCA-WC approach is a feature of the perturbative strong-coupling (hybridization)
expansion on which OCA-WC is based. In the correlated metallic
phase as well as in the vicinity of the Mott transition, the interplay
between various degrees of freedom and the approximations
inherent in the impurity solvers results in nontrivial effects on the spectral function. 
We have shown that as a consequence of the self-consistent
electron-phonon interaction, the electronic
charge fluctuations effectively reduce the vibrational frequency of the
phonons almost proportionally to $g^{2}/U$  in the weak electron-phonon coupling regime. We furthermore studied 
the low-energy physics of the system and determined a purely-electronic static interaction
which reproduces the low-energy spectral properties
of the Hubbard-Holstein model in the adiabatic
regime.\cite{Sangiovanni2005}

In an initially uncoupled system ($g=0$), we have switched on the electron-phonon
coupling to a moderate value and investigated the temporal
evolution of the system in various parameter regimes. We have
shown that in the correlated metallic phase of the uncoupled system, the initial dynamics produces a spectral-density reduction
of the Hubbard-bands and an enhancement of the quasi-particle peak which
continues, at most, until the phonon characteristic
time~($\pi/\omega_{\rm 0}$)
and is accompanied by an increase in the
double occupancy. These quasi-particles are later dressed by the
phonon cloud with the electron-mediated reduced frequency 
which results in a reduction of the
low-energy spectral density. For the Mott insulating initial phase
with a large el-el interaction, due to the very small
charge fluctuations as well as negligible thermal excitations, the
system does not show a dramatic redistribution of spectral weight. 
Close to the metal-to-insulator transition, where the gap size
is small, however, the initially uncoupled insulating state develops a
quasiparticle peak along a nonthermal trajectory. In this regime, spectral weight is transiently transferred
from the Hubbard bands to low energies and starts forming
a quasi-particle peak. The build-up of this peak is much more pronounced 
within SR-WC than in OCA-WC. We have discussed that this distinct response
is a result of the associated energy of the transferred
spectral densities in these two approximations. While within SR-WC
the spectral weight loss at the inner edge of Hubbard bands constructs the quasiparticle
peak, in OCA-WC the accumulated low-energy spectral densities are
mostly originating from the middle of the Hubbard bands.
By assessing the quasi-equilibrium condition
we have shown that the transient state is following 
a non-thermal pathway with distinguishable behavior at low and high energies.
We have also discussed that the quasi-particle weight (approximated via the integrations of the PES spectrum) suggest a static Hubbard interaction which
gradually decreases toward the correlated metallic phase. 
Questions
concerning the long-time thermalization of these induced non-thermal
metallic states, as well as their associated lifetimes, may be the subject
of future investigations. The overall investigation of the out-of-equilibrium dynamics revealed considerable discrepancies between the different methods, which illustrates the uncertainties associated with the use of the current state-of-the-art nonequilibrium impurity solvers.

Experiments on light-induced IMT transitions typically
observe the formation of a bad-metallic phase after ultra-fast laser excitations of Mott
insulators.\cite{Perfetti2006,Tobey2008,Hu2016,Kaiser2017,Iwai2003,Lysenko2007} 
Due to the energy injected by the pumping pulse, hot electron
carriers will be created and the role of the electron-lattice coupling,
amplified by the larger induced charge fluctuations, is mainly to cool
down these charge carries, which results in a slow reduction of the in-gap
density of states.\cite{Eckstein2013} This observation is very much consistent
with our presented picture in the large $U$ regime. Investigations of the
relaxation dynamics of heavy fermions, on the other hand, highlight the
importance of the low-energy physics in determining the 
thermalization timescale.\cite{Demsar2003} This is indeed one of our
main conclusions regarding the possibility of enhancing the metallic
tendencies in an insulating system with a small gap. But
whether the slow long time dynamics due to phonon dressing
is the dominating factor in addressing
the long electron-phonon relaxation time 
is an intriguing question which
requires extending our formalism to the study of Kondo-lattice type problems. 
It would also be very interesting to study multi-band systems to understand the
interplay between charge, orbital and phonon degrees of freedom in
inducing nontrivial metallic behaviors near the Mott transition.

\acknowledgments
We thank Y. Murakami for helpful discussions. The calculations have been performed on the PhysNET
cluster at the University of Hamburg, the REIMS cluster at the
Institute for Solid State Physics, and on the Beo04 cluster at the University of Fribourg.
Sh. S. is supported by the ImPACT Program of the Council for Science, Technology and
Innovation, Cabinet Office, Government of Japan (Grant No. 2015-PM12-05-01) from JST.
R\v{Z} acknowledges the support of the Slovenian Research Agency (ARRS)
under P1-0044 and J1-7259. DG and PW were supported by ERC Consolidator Grant 724103 and Swiss National Science Foundation Grant 200021-165539. The Flatiron Institute is a division of the Simons Foundation.

\newpage

\appendix

\section{Equilibrium spectral functions}\label{appendix}
\begin{figure}[t]
\includegraphics{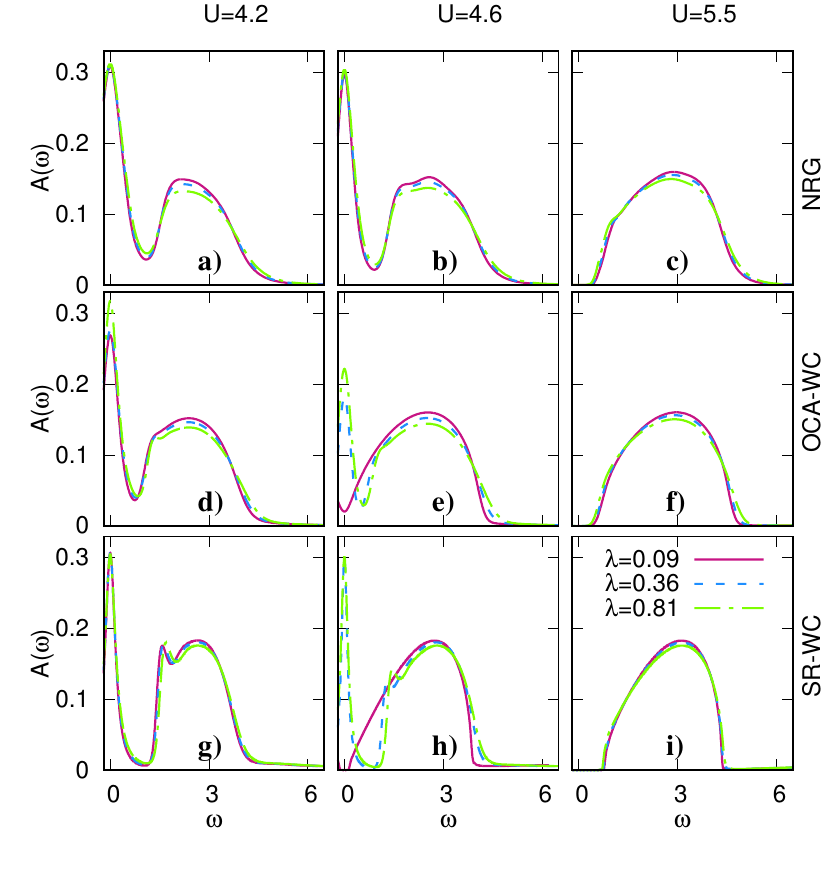}
\caption{ Equilibrium spectral function~$A(\omega)$ obtained from OCA-WC, NRG, and SR-WC for 
$\lambda \in \{ 0.09~\text{(red lines), } 0.36~\text{(blue lines), } 0.81~\text{(green lines)} \}$,
$\omega_{0}=0.2$ and $U \in \{ 4.2, 4.6 ,5.5 \}$.
Panels on the same row are computed using the indicated approximation. 
The vertically aligned panels describe systems at a fixed Hubbard interaction.
}
\label{Fig:Spect3Uw02}
\end{figure}

\begin{figure}[t]
\includegraphics{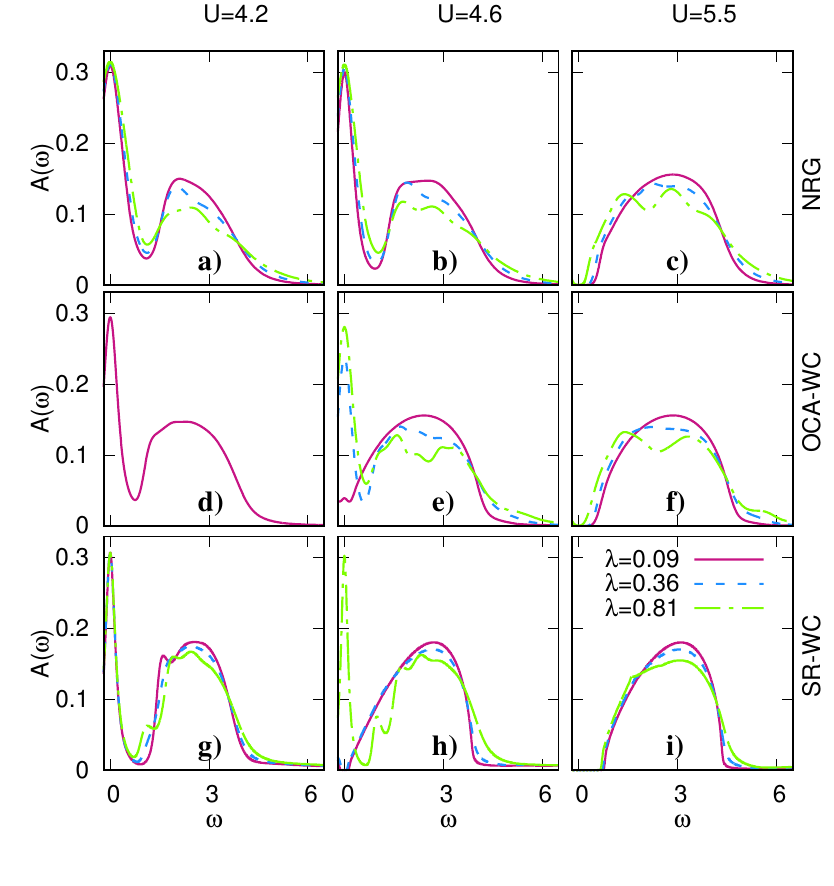}
\caption{ Equilibrium spectral function~$A(\omega)$ obtained from OCA-WC, NRG, and SR-WC for 
$\lambda \in \{ 0.09~\text{(red lines), } 0.36~\text{(blue lines), } 0.81~\text{(green lines)} \}$,
$\omega_{0}=1.0$ and $U \in \{ 4.2, 4.6 ,5.5 \}$.
Panels on the same row are computed within the mentioned approximation. 
The vertically aligned panels describe systems at a fixed electron-phonon coupling.
Missing OCA data indicate that the solutions cannot be converged.
}
\label{Fig:Spect3Uw1}
\end{figure}

In order to illustrate the evolution of the spectral functions for increasing electron-phonon
interaction strengths within a given approximation we rearranged the
data from Sec.~\ref{Sec:Spectral}. The adiabatic cases for
$\omega_0=0.2$ are presented in
Fig.~\ref{Fig:Spect3Uw02} and those for the high phonon frequency
$\omega_0=1.0$ in Fig.~\ref{Fig:Spect3Uw1}. The 
spectral features are discussed in Sec.~\ref{Sec:Spectral}.

\section{Renormalized phonon frequency}\label{Supp:renormphonon}

To determine the relationship between the renormalized phonon frequency
and other physical parameters of the system,
we will consider the SR-based formalism.
Within DMFT, the major contribution of the phonon softening for local
electron-phonon interactions is conveyed by space-local terms. 
The associated effective action can be written as 
\begin{align}
 {\cal S}_{\rm nph} = \int_{\cal C} {\rm d}t 
 & \left[
 X_{\rm ph}(t) {\cal D}^{-1}_{0}(t,t) X_{\rm ph}(t)
 -\sqrt{2} g(t) n(t) X_{\rm ph}(t) \right. \nonumber  \\
  & \left.
 +U \sum\limits_{\sigma} n_{\sigma} n_{\overline{\sigma}}
 \right],
\end{align}
where $\cal C$ denotes the Keldysh contour, $\sigma$ stands for the spin index, 
$X_{\rm ph}$ is the phonon displacement operator given by $X_{\rm ph}= (b+b^{\dagger})/\sqrt{2}$, and 
${\cal D}^{-1}_{0}$ is the noninteracting phonon propagator defined as
${\cal D}_{0}= - (\partial_{t}^{2} + \omega_{0}^{2})/2\omega_{0}$.
Here we have dropped the site indices for simplicity.
Within the slave-rotor decomposition~\cite{Florens2002,Sayyad2016} the above action can be rewritten as
\begin{align}
 {\cal S}_{\rm L \theta ph}=\int_{\cal C} {\rm d}t 
  &\left[ 
 -U L^{2}(t)  
 +X_{\rm ph}(t) {\cal D}^{-1}_{0}(t,t) X_{\rm ph}(t)
  \right. \nonumber  \\
 & \left. 
 -\sqrt{2} g(t) L(t) X_{\rm ph}(t)
 + \eta L(t)
 +L(t) \partial_{t} \theta
 \right],
\end{align}
where $\theta$ is the canonical angle of the rotor angular momentum~($L$), and 
$\eta$ is the Lagrange multiplier to maintain the charge-conservation. 
Performing the functional integral over the rotor angular momentum yields 
\begin{align}
 &{\cal S}_{\rm \theta ph}=\int_{\cal C} {\rm d}t 
  \Big[ 
  X_{\rm ph}(t) {\cal D}^{-1}_{0}(t,t) X_{\rm ph}(t)
   \nonumber  \\
 & 
  +\frac{1}{2}\big( \partial_{t} \theta  + \eta - \sqrt{2} g(t) X_{\rm ph}(t) \big)
  \frac{1}{U}
  \big( \partial_{t} \theta  + \eta - \sqrt{2} g(t) X_{\rm ph}(t) \big)
  \Big].
\end{align}
Incorporating the quadratic terms in $X_{\rm ph}$ from the second term of ${\cal S}_{\rm \theta ph}$
into its first term, we would obtain a renormalized phonon Green's function satisfying
\begin{align}
 {\cal D}_{\rm r}&=- \frac{\partial_{t}^{2} + \omega_{0}^{2}}{2\omega_{0}} + \frac{g^{2}}{U},\\
 &\stackrel{ \frac{2g^{2}}{U}\ll\omega_{0}}{=}- \frac{\partial_{t}^{2} 
 + \omega_{\rm r}^{2}}{2\omega_{\rm r}} .
\end{align}
In the limit where ${2g^{2}}/{U}\ll\omega_{\rm 0}$, we therefore 
estimate the phonon softening as
\begin{equation}
 \omega_{\rm r}\approx \omega_{0} - \frac{g^{2}}{U}.
\end{equation}
Away from this regime, the associated charge fluctuations induced by $\partial_{t} \theta$ produce nonlinear effects.

\begin{figure}[t]
\includegraphics{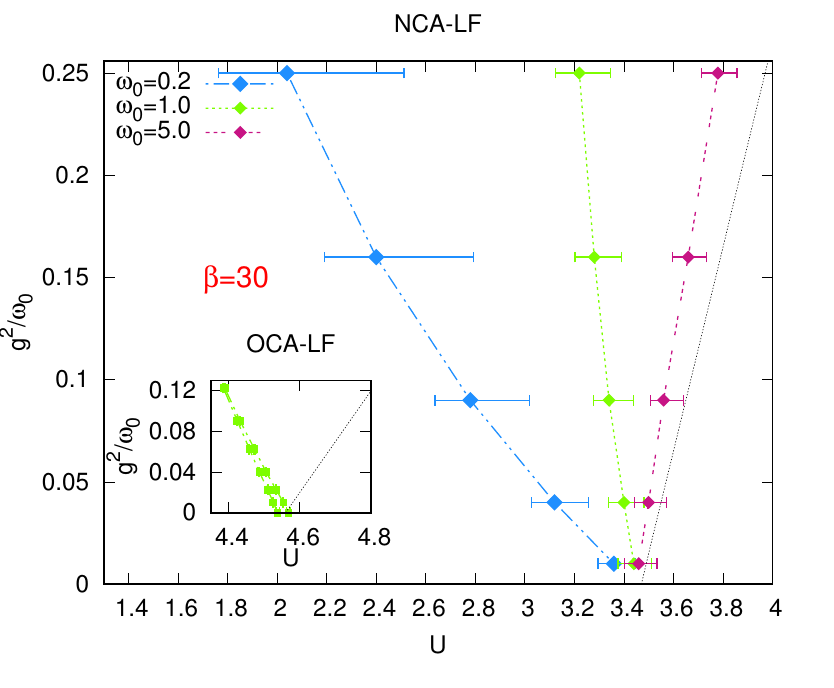}\\[-5mm]
\caption{\label{fig:LF_phase_diagram} (Color online)
Metal to Mott insulator crossover in the non-crossing approximation with the Lang-Firsov transformation (NCA-LF)  for $\beta = 30$ and $\omega_0 \in \{ 0.2, 1, 5 \}$.
Inset: Metal to Mott insulator phase boundaries in the one crossing approximation with Lang-Firsov transformation (OCA-LF) at $\beta = 30$ and $\omega_0 = 0.2$.
The phase boundary in the large phonon-frequency limit, $U_c = U_c(g\!=\!0) + \frac{2g^2}{\omega_0}$, is also shown  (dotted black lines).
}
\end{figure}

\section{Strong coupling expansion combined with the Lang-Firsov transformation}\label{Supp:LF}

In the hybridization expansion, an alternative to the weak coupling expansion in the electron-phonon coupling (e.g.\ NCA-WC and OCA-WC), is to apply a Lang-Firsov (LF) \cite{Lang:1963aa} decoupling of the electron-phonon interaction, i.e., a transformation to polaron operators. 
In combination with the hybridization expansion this transformation enables numerically exact simulations of the Hubbard-Holstein model in DMFT,\cite{Werner2007} using continuous time quantum Monte Carlo (CTQMC).\cite{Werner:2006qy}  Out of equilibrium, it has been used in combination with NCA and OCA to study doublon relaxation in the single band Hubbard-Holstein model.\cite{Werner2013}

We have implemented NCA-LF and OCA-LF in the simplest approximation, described in detail in Ref.~\onlinecite{Werner2013}, which effectively amounts to dress each pair of fermionic creation-annihilation operators in the perturbation theory with an additional bosonic factor, see Eq.\ (30) in Ref.~\onlinecite{Werner2013}, and a phonon induced shift $U \rightarrow U - 2g^2/\omega_0$ of the Hubbard interaction.
The resulting approximation is different from the weak coupling expansion in the electron-phonon coupling $g$, since it captures the Mott to bipolaronic transition at large $g$.\cite{Werner2013} However, 

as we will show, it gives qualitatively correct results only in the large-$U$ and large-$\omega_0$ regime.

The Monte-Carlo sampling of the bare strong coupling expansion is exact and accounts for all bosonic contributions\cite{Werner2007} by connecting all fermionic operators in the partition function expansion with the bosonic ``weight'' factors generated by the Lang-Firsov transformation.
The dressed strong coupling approach, however, performs an expansion where the atomic propagator is dressed with low order self-energy expansions in the hybridization function, re-summed to infinite order using the Dyson equation.
The bosonic weight factors of operator pairs are only accounted for within each strong-coupling self-energy diagram.
This is an approximation since the bosonic weight factors associated with pairings of fermion operators \emph{between} self-energy insertions in the Dyson equation are neglected. This is also the case in the diagrams for the single-particle Green's function.
For this reason, NCA-LF and OCA-LF are accurate only if the bosonic weight factors decay fast, or oscillate rapidly, which is the case in the limit of large $\omega_0$. 

To demonstrate the limitations of NCA-LF and OCA-LF we map out the metal-insulator phase boundary at low $g$, see Fig.~\ref{fig:LF_phase_diagram}. In NCA-LF the transition is a crossover, whose center $U_c$ is determined here by the maxima of the second order derivative in the double occupancy, i.e.\ $U_c = \max_U |\partial^2_U \langle \hat{n}_\uparrow \hat{n}_\downarrow \rangle|$, while the extent of the crossover region is determined by the corresponding width at half maximum.
Small phonon frequencies ($\omega_0 = 0.2$) yields a decreasing $U_c$ with increasing $g$, while at large frequencies $U_c$ approaches the expected high frequency limit, $U_c \approx U_c(g\!\!=\!\!0) + \frac{2g^2}{\omega_0}$, having the opposite slope in $g$. We note that the reduction in $U_c$ upon increasing $g$ at low $\omega_0$ is qualitatively different from the exact Monte Carlo result in Fig.~\ref{Fig:phasediagram}.

While all NCA based approximations under-estimate $U_c(g=0)$, this is improved when using OCA. The hysteresis region of OCA-LF at $\omega_0 = 0.2$ is shown in the inset of Fig.\ \ref{fig:LF_phase_diagram}. However, while $U_c(g=0)$ is closer the CTQMC result, the $U_c$ dependence on $g$ remains qualitatively wrong as for NCA-LF.
We conclude that LF based second order strong coupling approximation (OCA-LF) does not qualitatively capture the metal to Mott phase boundary in the weak electron-phonon coupling and small-$\omega_0$ regime.

Although the behavior near the MIT 
is not correctly described in OCA-LF we find that the results of this method are qualitatively correct in the strong coupling regime and for large enough $\omega_0$. Since it is the only real-time non-equilibrium approach that captures the Mott to bipolaronic transition, a comparison of its equilibrium spectral function with NRG is of interest.
\begin{figure}[t]\label{Fig:SC1}
\includegraphics{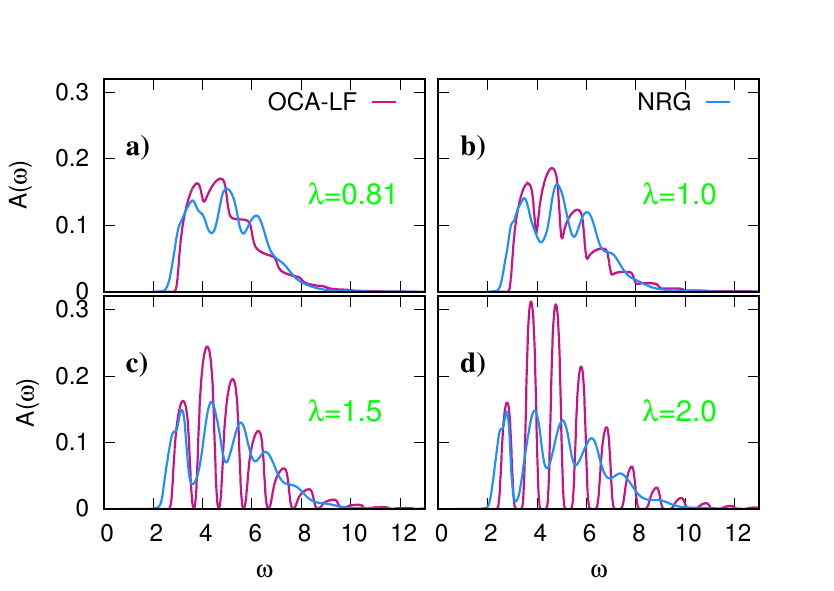}\\[-3mm]
\caption{\label{fig:NRG_OCA_LF_omega_1_0} (Color online) Spectral functions from OCA-LF and NRG averaged over six different discretization parameters, namely $\Lambda \in \{ 1.8,1.9,2.0,2.1,2.2,2.3\}$ and reduced broadening parameter $\alpha=0.05$, at $U=10$, $\beta = 30$, and $\omega_0 = 1$ for varying electron phonon coupling strength $g^2/\omega_0 \in \{ 0.81, 1, 1.5, 2 \}$.  
}
\end{figure}
\begin{figure}
\includegraphics{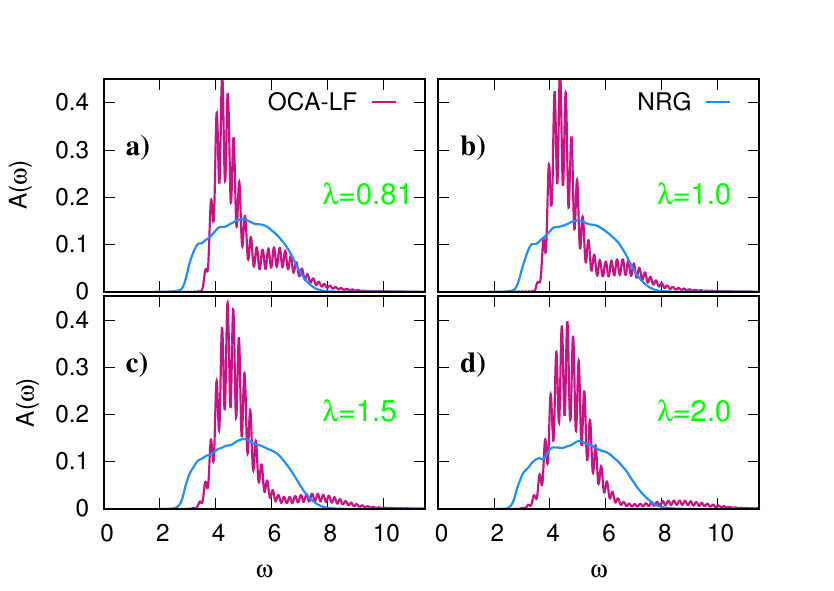}\\[-3mm]
\caption{\label{fig:NRG_OCA_LF_omega_0_2} (Color online) Spectral functions from OCA-LF and NRG averaged over six different discretization parameters, namely $\Lambda \in \{ 1.8,1.9,2.0,2.1,2.2,2.3\}$ and reduced broadening parameter $\alpha=0.05$, at $U=10$, $\beta = 30$, and $\omega_0 = 0.2$ for varying electron phonon coupling strength $g^2/\omega_0 \in \{ 0.81, 1, 1.5, 2 \}$.}
\label{Fig:SpectLFNRGw02}
\end{figure}
Due to the strong electron-phonon interaction, sharp polaronic features are expected in the spectrum, which can be smeared out by NRG broadening of the raw spectra. Therefore, in the following, we present spectra in Fig.~\ref{fig:NRG_OCA_LF_omega_1_0} and \ref{Fig:SpectLFNRGw02} for the reduced broadening $\alpha=0.05$ averaged over six different discretization parameters $\Lambda\in\{1.8,1.9,2.0,2.1,2.2,2.3\}$, in order to distinguish sharp features.

\begin{figure}
\includegraphics{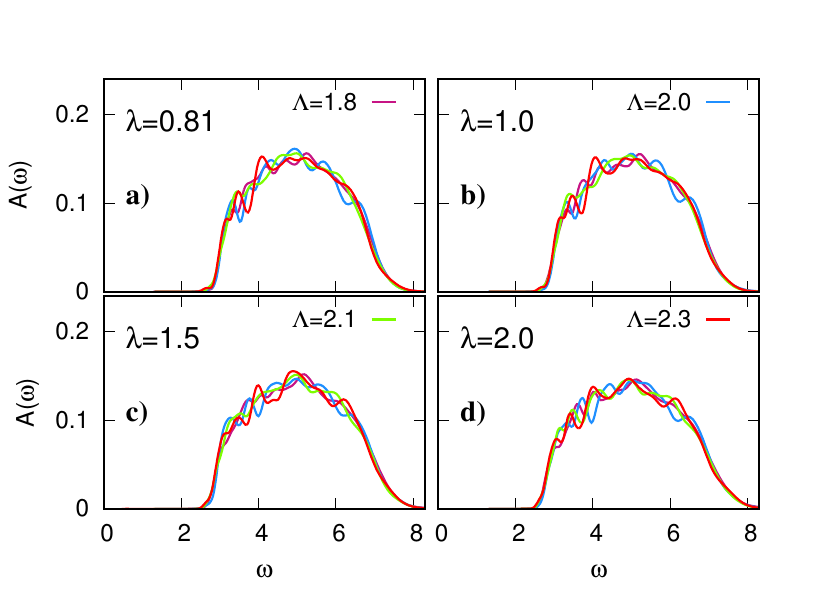}\\[-3mm]
\caption{\label{fig:NRG_omega_1} (Color online) Spectral functions from NRG at four different discretization parameters, namely $\Lambda\in \{ 1.8, 2.0,2.1,2.3 \}$ and reduced broadening parameter $\alpha=0.05$, at $U=10$, $\beta = 30$, and $\omega_0 = 0.2$ for varying electron phonon coupling strength $g^2/\omega_0 \in \{ 0.81, 1, 1.5, 2 \}$.}
\label{Fig:SpectNRGw02}
\end{figure}

At $U = 10$ and $\beta = 30$ and large phonon frequency $\omega_0 = 1$, the NRG spectral function shows a fine structure of the Hubbard band and a broad tail at high energies, in qualitative agreement with the OCA-LF result. For weaker electron-phonon interaction the separation between the peaks is larger than the phonon frequency $\omega_0 = 1$, see for instance $\lambda=0.81,1.0$ in Fig.~\ref{fig:NRG_OCA_LF_omega_1_0}, while for the strongest electron-phonon interaction the separation between the peaks is clearly given by the phonon quanta $\omega_0$. 
Reducing the phonon frequency to $\omega_0 = 0.2$ yields stronger discrepancies between NRG and OCA-LF, see Fig.~\ref{fig:NRG_OCA_LF_omega_0_2}. Phonon peaks can be observed in the OCA-LF spectral function, while in the NRG result they are completely washed out. For weak broadening the position of the peak depends on the discretization parameter $\Lambda$ and it is hard to obtain discretization-parameter independent results, see also Fig.~\ref{Fig:SpectNRGw02} for the comparison of the spectra for different discretization parameters $\Lambda$. 
The main reason for the discrepancy is, however, the expected inaccuracy of the OCA-LF method in the adiabatic regime, where the bosonic weight factors are slowly-varying so that the approximations inherent in the perturbative approach become more severe, independent of the value of $U$.  

\bibliography{report}

\begin{thebibliography}{69}%
\makeatletter
\providecommand \@ifxundefined [1]{%
 \@ifx{#1\undefined}
}%
\providecommand \@ifnum [1]{%
 \ifnum #1\expandafter \@firstoftwo
 \else \expandafter \@secondoftwo
 \fi
}%
\providecommand \@ifx [1]{%
 \ifx #1\expandafter \@firstoftwo
 \else \expandafter \@secondoftwo
 \fi
}%
\providecommand \natexlab [1]{#1}%
\providecommand \enquote  [1]{``#1''}%
\providecommand \bibnamefont  [1]{#1}%
\providecommand \bibfnamefont [1]{#1}%
\providecommand \citenamefont [1]{#1}%
\providecommand \href@noop [0]{\@secondoftwo}%
\providecommand \href [0]{\begingroup \@sanitize@url \@href}%
\providecommand \@href[1]{\@@startlink{#1}\@@href}%
\providecommand \@@href[1]{\endgroup#1\@@endlink}%
\providecommand \@sanitize@url [0]{\catcode `\\12\catcode `\$12\catcode
  `\&12\catcode `\#12\catcode `\^12\catcode `\_12\catcode `\%12\relax}%
\providecommand \@@startlink[1]{}%
\providecommand \@@endlink[0]{}%
\providecommand \url  [0]{\begingroup\@sanitize@url \@url }%
\providecommand \@url [1]{\endgroup\@href {#1}{\urlprefix }}%
\providecommand \urlprefix  [0]{URL }%
\providecommand \Eprint [0]{\href }%
\providecommand \doibase [0]{http://dx.doi.org/}%
\providecommand \selectlanguage [0]{\@gobble}%
\providecommand \bibinfo  [0]{\@secondoftwo}%
\providecommand \bibfield  [0]{\@secondoftwo}%
\providecommand \translation [1]{[#1]}%
\providecommand \BibitemOpen [0]{}%
\providecommand \bibitemStop [0]{}%
\providecommand \bibitemNoStop [0]{.\EOS\space}%
\providecommand \EOS [0]{\spacefactor3000\relax}%
\providecommand \BibitemShut  [1]{\csname bibitem#1\endcsname}%
\let\auto@bib@innerbib\@empty
\bibitem [{\citenamefont {Morin}(1959)}]{Morin1959}%
  \BibitemOpen
  \bibfield  {author} {\bibinfo {author} {\bibfnamefont {F.~J.}\ \bibnamefont
  {Morin}},\ }\href {\doibase 10.1103/PhysRevLett.3.34} {\bibfield  {journal}
  {\bibinfo  {journal} {Phys. Rev. Lett.}\ }\textbf {\bibinfo {volume} {3}},\
  \bibinfo {pages} {34} (\bibinfo {year} {1959})}\BibitemShut {NoStop}%
\bibitem [{\citenamefont {Qiu}\ \emph {et~al.}(2017)\citenamefont {Qiu},
  \citenamefont {Bousquet},\ and\ \citenamefont {Cano}}]{Qiu2017}%
  \BibitemOpen
  \bibfield  {author} {\bibinfo {author} {\bibfnamefont {R.}~\bibnamefont
  {Qiu}}, \bibinfo {author} {\bibfnamefont {E.}~\bibnamefont {Bousquet}}, \
  and\ \bibinfo {author} {\bibfnamefont {A.}~\bibnamefont {Cano}},\ }\href
  {http://stacks.iop.org/0953-8984/29/i=30/a=305801} {\bibfield  {journal}
  {\bibinfo  {journal} {Journal of Physics: Condensed Matter}\ }\textbf
  {\bibinfo {volume} {29}},\ \bibinfo {pages} {305801} (\bibinfo {year}
  {2017})}\BibitemShut {NoStop}%
\bibitem [{\citenamefont {Gavriliuk}\ \emph {et~al.}(2012)\citenamefont
  {Gavriliuk}, \citenamefont {Trojan},\ and\ \citenamefont
  {Struzhkin}}]{Gavriliuk2012}%
  \BibitemOpen
  \bibfield  {author} {\bibinfo {author} {\bibfnamefont {A.~G.}\ \bibnamefont
  {Gavriliuk}}, \bibinfo {author} {\bibfnamefont {I.~A.}\ \bibnamefont
  {Trojan}}, \ and\ \bibinfo {author} {\bibfnamefont {V.~V.}\ \bibnamefont
  {Struzhkin}},\ }\href {\doibase 10.1103/PhysRevLett.109.086402} {\bibfield
  {journal} {\bibinfo  {journal} {Phys. Rev. Lett.}\ }\textbf {\bibinfo
  {volume} {109}},\ \bibinfo {pages} {086402} (\bibinfo {year}
  {2012})}\BibitemShut {NoStop}%
\bibitem [{\citenamefont {Perfetti}\ \emph {et~al.}(2006)\citenamefont
  {Perfetti}, \citenamefont {Loukakos}, \citenamefont {Lisowski}, \citenamefont
  {Bovensiepen}, \citenamefont {Berger}, \citenamefont {Biermann},
  \citenamefont {Cornaglia}, \citenamefont {Georges},\ and\ \citenamefont
  {Wolf}}]{Perfetti2006}%
  \BibitemOpen
  \bibfield  {author} {\bibinfo {author} {\bibfnamefont {L.}~\bibnamefont
  {Perfetti}}, \bibinfo {author} {\bibfnamefont {P.~A.}\ \bibnamefont
  {Loukakos}}, \bibinfo {author} {\bibfnamefont {M.}~\bibnamefont {Lisowski}},
  \bibinfo {author} {\bibfnamefont {U.}~\bibnamefont {Bovensiepen}}, \bibinfo
  {author} {\bibfnamefont {H.}~\bibnamefont {Berger}}, \bibinfo {author}
  {\bibfnamefont {S.}~\bibnamefont {Biermann}}, \bibinfo {author}
  {\bibfnamefont {P.~S.}\ \bibnamefont {Cornaglia}}, \bibinfo {author}
  {\bibfnamefont {A.}~\bibnamefont {Georges}}, \ and\ \bibinfo {author}
  {\bibfnamefont {M.}~\bibnamefont {Wolf}},\ }\href {\doibase
  10.1103/PhysRevLett.97.067402} {\bibfield  {journal} {\bibinfo  {journal}
  {Phys. Rev. Lett.}\ }\textbf {\bibinfo {volume} {97}},\ \bibinfo {pages}
  {067402} (\bibinfo {year} {2006})}\BibitemShut {NoStop}%
\bibitem [{\citenamefont {Tobey}\ \emph {et~al.}(2008)\citenamefont {Tobey},
  \citenamefont {Prabhakaran}, \citenamefont {Boothroyd},\ and\ \citenamefont
  {Cavalleri}}]{Tobey2008}%
  \BibitemOpen
  \bibfield  {author} {\bibinfo {author} {\bibfnamefont {R.~I.}\ \bibnamefont
  {Tobey}}, \bibinfo {author} {\bibfnamefont {D.}~\bibnamefont {Prabhakaran}},
  \bibinfo {author} {\bibfnamefont {A.~T.}\ \bibnamefont {Boothroyd}}, \ and\
  \bibinfo {author} {\bibfnamefont {A.}~\bibnamefont {Cavalleri}},\ }\href
  {\doibase 10.1103/PhysRevLett.101.197404} {\bibfield  {journal} {\bibinfo
  {journal} {Phys. Rev. Lett.}\ }\textbf {\bibinfo {volume} {101}},\ \bibinfo
  {pages} {197404} (\bibinfo {year} {2008})}\BibitemShut {NoStop}%
\bibitem [{\citenamefont {Hu}\ \emph {et~al.}(2016)\citenamefont {Hu},
  \citenamefont {Catalano}, \citenamefont {Gibert}, \citenamefont {Triscone},\
  and\ \citenamefont {Cavalleri}}]{Hu2016}%
  \BibitemOpen
  \bibfield  {author} {\bibinfo {author} {\bibfnamefont {W.}~\bibnamefont
  {Hu}}, \bibinfo {author} {\bibfnamefont {S.}~\bibnamefont {Catalano}},
  \bibinfo {author} {\bibfnamefont {M.}~\bibnamefont {Gibert}}, \bibinfo
  {author} {\bibfnamefont {J.-M.}\ \bibnamefont {Triscone}}, \ and\ \bibinfo
  {author} {\bibfnamefont {A.}~\bibnamefont {Cavalleri}},\ }\href {\doibase
  10.1103/PhysRevB.93.161107} {\bibfield  {journal} {\bibinfo  {journal} {Phys.
  Rev. B}\ }\textbf {\bibinfo {volume} {93}},\ \bibinfo {pages} {161107}
  (\bibinfo {year} {2016})}\BibitemShut {NoStop}%
\bibitem [{\citenamefont {Kaiser}(2017)}]{Kaiser2017}%
  \BibitemOpen
  \bibfield  {author} {\bibinfo {author} {\bibfnamefont {S.}~\bibnamefont
  {Kaiser}},\ }\href {http://stacks.iop.org/1402-4896/92/i=10/a=103001}
  {\bibfield  {journal} {\bibinfo  {journal} {Physica Scripta}\ }\textbf
  {\bibinfo {volume} {92}},\ \bibinfo {pages} {103001} (\bibinfo {year}
  {2017})}\BibitemShut {NoStop}%
\bibitem [{\citenamefont {Eckstein}\ and\ \citenamefont
  {Werner}(2013)}]{Eckstein2013}%
  \BibitemOpen
  \bibfield  {author} {\bibinfo {author} {\bibfnamefont {M.}~\bibnamefont
  {Eckstein}}\ and\ \bibinfo {author} {\bibfnamefont {P.}~\bibnamefont
  {Werner}},\ }\href {\doibase 10.1103/PhysRevLett.110.126401} {\bibfield
  {journal} {\bibinfo  {journal} {Phys. Rev. Lett.}\ }\textbf {\bibinfo
  {volume} {110}},\ \bibinfo {pages} {126401} (\bibinfo {year}
  {2013})}\BibitemShut {NoStop}%
\bibitem [{\citenamefont {Sayyad}\ and\ \citenamefont
  {Eckstein}(2016)}]{Sayyad2016}%
  \BibitemOpen
  \bibfield  {author} {\bibinfo {author} {\bibfnamefont {S.}~\bibnamefont
  {Sayyad}}\ and\ \bibinfo {author} {\bibfnamefont {M.}~\bibnamefont
  {Eckstein}},\ }\href {\doibase 10.1103/PhysRevLett.117.096403} {\bibfield
  {journal} {\bibinfo  {journal} {Phys. Rev. Lett.}\ }\textbf {\bibinfo
  {volume} {117}},\ \bibinfo {pages} {96403} (\bibinfo {year}
  {2016})}\BibitemShut {NoStop}%
\bibitem [{\citenamefont {Werner}\ and\ \citenamefont
  {Millis}(2007)}]{Werner2007}%
  \BibitemOpen
  \bibfield  {author} {\bibinfo {author} {\bibfnamefont {P.}~\bibnamefont
  {Werner}}\ and\ \bibinfo {author} {\bibfnamefont {A.}~\bibnamefont
  {Millis}},\ }\href {\doibase 10.1103/PhysRevLett.99.146404} {\bibfield
  {journal} {\bibinfo  {journal} {Physical Review Letters}\ }\textbf {\bibinfo
  {volume} {99}},\ \bibinfo {pages} {146404} (\bibinfo {year}
  {2007})}\BibitemShut {NoStop}%
\bibitem [{\citenamefont {Koller}\ \emph
  {et~al.}(2004{\natexlab{a}})\citenamefont {Koller}, \citenamefont {Meyer},
  \citenamefont {Ono},\ and\ \citenamefont {Hewson}}]{Koller2004}%
  \BibitemOpen
  \bibfield  {author} {\bibinfo {author} {\bibfnamefont {W.}~\bibnamefont
  {Koller}}, \bibinfo {author} {\bibfnamefont {D.}~\bibnamefont {Meyer}},
  \bibinfo {author} {\bibfnamefont {Y.}~\bibnamefont {Ono}}, \ and\ \bibinfo
  {author} {\bibfnamefont {A.~C.}\ \bibnamefont {Hewson}},\ }\href {\doibase
  10.1209/epl/i2003-10228-6} {\bibfield  {journal} {\bibinfo  {journal}
  {Europhysics Letters}\ }\textbf {\bibinfo {volume} {66}},\ \bibinfo {pages}
  {559} (\bibinfo {year} {2004}{\natexlab{a}})}\BibitemShut {NoStop}%
\bibitem [{\citenamefont {Werner}\ and\ \citenamefont
  {Millis}(2010)}]{Werner2010}%
  \BibitemOpen
  \bibfield  {author} {\bibinfo {author} {\bibfnamefont {P.}~\bibnamefont
  {Werner}}\ and\ \bibinfo {author} {\bibfnamefont {A.~J.}\ \bibnamefont
  {Millis}},\ }\href {\doibase 10.1103/PhysRevLett.104.146401} {\bibfield
  {journal} {\bibinfo  {journal} {Phys. Rev. Lett.}\ }\textbf {\bibinfo
  {volume} {104}},\ \bibinfo {pages} {146401} (\bibinfo {year}
  {2010})}\BibitemShut {NoStop}%
\bibitem [{\citenamefont {Gole\ifmmode~\check{z}\else \v{z}\fi{}}\ \emph
  {et~al.}(2015)\citenamefont {Gole\ifmmode~\check{z}\else \v{z}\fi{}},
  \citenamefont {Eckstein},\ and\ \citenamefont {Werner}}]{Golez2015}%
  \BibitemOpen
  \bibfield  {author} {\bibinfo {author} {\bibfnamefont {D.}~\bibnamefont
  {Gole\ifmmode~\check{z}\else \v{z}\fi{}}}, \bibinfo {author} {\bibfnamefont
  {M.}~\bibnamefont {Eckstein}}, \ and\ \bibinfo {author} {\bibfnamefont
  {P.}~\bibnamefont {Werner}},\ }\href {\doibase 10.1103/PhysRevB.92.195123}
  {\bibfield  {journal} {\bibinfo  {journal} {Phys. Rev. B}\ }\textbf {\bibinfo
  {volume} {92}},\ \bibinfo {pages} {195123} (\bibinfo {year}
  {2015})}\BibitemShut {NoStop}%
\bibitem [{\citenamefont {Jeon}\ \emph {et~al.}(2004)\citenamefont {Jeon},
  \citenamefont {Park}, \citenamefont {Han}, \citenamefont {Lee},\ and\
  \citenamefont {Choi}}]{Jeon2004}%
  \BibitemOpen
  \bibfield  {author} {\bibinfo {author} {\bibfnamefont {G.~S.}\ \bibnamefont
  {Jeon}}, \bibinfo {author} {\bibfnamefont {T.-H.}\ \bibnamefont {Park}},
  \bibinfo {author} {\bibfnamefont {J.~H.}\ \bibnamefont {Han}}, \bibinfo
  {author} {\bibfnamefont {H.~C.}\ \bibnamefont {Lee}}, \ and\ \bibinfo
  {author} {\bibfnamefont {H.-Y.}\ \bibnamefont {Choi}},\ }\href {\doibase
  10.1103/PhysRevB.70.125114} {\bibfield  {journal} {\bibinfo  {journal} {Phys.
  Rev. B}\ }\textbf {\bibinfo {volume} {70}},\ \bibinfo {pages} {125114}
  (\bibinfo {year} {2004})}\BibitemShut {NoStop}%
\bibitem [{\citenamefont {Sangiovanni}\ \emph {et~al.}(2005)\citenamefont
  {Sangiovanni}, \citenamefont {Capone}, \citenamefont {Castellani},\ and\
  \citenamefont {Grilli}}]{Sangiovanni2005}%
  \BibitemOpen
  \bibfield  {author} {\bibinfo {author} {\bibfnamefont {G.}~\bibnamefont
  {Sangiovanni}}, \bibinfo {author} {\bibfnamefont {M.}~\bibnamefont {Capone}},
  \bibinfo {author} {\bibfnamefont {C.}~\bibnamefont {Castellani}}, \ and\
  \bibinfo {author} {\bibfnamefont {M.}~\bibnamefont {Grilli}},\ }\href
  {\doibase 10.1103/PhysRevLett.94.026401} {\bibfield  {journal} {\bibinfo
  {journal} {Phys. Rev. Lett.}\ }\textbf {\bibinfo {volume} {94}},\ \bibinfo
  {pages} {026401} (\bibinfo {year} {2005})}\BibitemShut {NoStop}%
\bibitem [{\citenamefont {Sangiovanni}\ \emph {et~al.}(2006)\citenamefont
  {Sangiovanni}, \citenamefont {Capone},\ and\ \citenamefont
  {Castellani}}]{Sangiovanni2006}%
  \BibitemOpen
  \bibfield  {author} {\bibinfo {author} {\bibfnamefont {G.}~\bibnamefont
  {Sangiovanni}}, \bibinfo {author} {\bibfnamefont {M.}~\bibnamefont {Capone}},
  \ and\ \bibinfo {author} {\bibfnamefont {C.}~\bibnamefont {Castellani}},\
  }\href {\doibase 10.1103/PhysRevB.73.165123} {\bibfield  {journal} {\bibinfo
  {journal} {Phys. Rev. B}\ }\textbf {\bibinfo {volume} {73}},\ \bibinfo
  {pages} {165123} (\bibinfo {year} {2006})}\BibitemShut {NoStop}%
\bibitem [{\citenamefont {Kapcia}\ \emph {et~al.}(2017)\citenamefont {Kapcia},
  \citenamefont {Robaszkiewicz}, \citenamefont {Capone},\ and\ \citenamefont
  {Amaricci}}]{Kapcia2017}%
  \BibitemOpen
  \bibfield  {author} {\bibinfo {author} {\bibfnamefont {K.~J.}\ \bibnamefont
  {Kapcia}}, \bibinfo {author} {\bibfnamefont {S.}~\bibnamefont
  {Robaszkiewicz}}, \bibinfo {author} {\bibfnamefont {M.}~\bibnamefont
  {Capone}}, \ and\ \bibinfo {author} {\bibfnamefont {A.}~\bibnamefont
  {Amaricci}},\ }\href {\doibase 10.1103/PhysRevB.95.125112} {\bibfield
  {journal} {\bibinfo  {journal} {Phys. Rev. B}\ }\textbf {\bibinfo {volume}
  {95}},\ \bibinfo {pages} {125112} (\bibinfo {year} {2017})}\BibitemShut
  {NoStop}%
\bibitem [{\citenamefont {Sch\"uler}\ \emph {et~al.}(2018)\citenamefont
  {Sch\"uler}, \citenamefont {van Loon}, \citenamefont {Katsnelson},\ and\
  \citenamefont {Wehling}}]{Schuler2017}%
  \BibitemOpen
  \bibfield  {author} {\bibinfo {author} {\bibfnamefont {M.}~\bibnamefont
  {Sch\"uler}}, \bibinfo {author} {\bibfnamefont {E.~G. C.~P.}\ \bibnamefont
  {van Loon}}, \bibinfo {author} {\bibfnamefont {M.~I.}\ \bibnamefont
  {Katsnelson}}, \ and\ \bibinfo {author} {\bibfnamefont {T.~O.}\ \bibnamefont
  {Wehling}},\ }\href {\doibase 10.1103/PhysRevB.97.165135} {\bibfield
  {journal} {\bibinfo  {journal} {Phys. Rev. B}\ }\textbf {\bibinfo {volume}
  {97}},\ \bibinfo {pages} {165135} (\bibinfo {year} {2018})}\BibitemShut
  {NoStop}%
\bibitem [{\citenamefont {Gao}\ and\ \citenamefont {Wang}(2009)}]{Gao2009}%
  \BibitemOpen
  \bibfield  {author} {\bibinfo {author} {\bibfnamefont {J.}~\bibnamefont
  {Gao}}\ and\ \bibinfo {author} {\bibfnamefont {J.}~\bibnamefont {Wang}},\
  }\href {\doibase 10.1088/0953-8984/21/48/485702} {\bibfield  {journal}
  {\bibinfo  {journal} {Journal of Physics: Condensed Matter}\ }\textbf
  {\bibinfo {volume} {21}},\ \bibinfo {pages} {485702} (\bibinfo {year}
  {2009})}\BibitemShut {NoStop}%
\bibitem [{\citenamefont {Babadi}\ \emph {et~al.}(2017)\citenamefont {Babadi},
  \citenamefont {Knap}, \citenamefont {Martin}, \citenamefont {Refael},\ and\
  \citenamefont {Demler}}]{babadi2017}%
  \BibitemOpen
  \bibfield  {author} {\bibinfo {author} {\bibfnamefont {M.}~\bibnamefont
  {Babadi}}, \bibinfo {author} {\bibfnamefont {M.}~\bibnamefont {Knap}},
  \bibinfo {author} {\bibfnamefont {I.}~\bibnamefont {Martin}}, \bibinfo
  {author} {\bibfnamefont {G.}~\bibnamefont {Refael}}, \ and\ \bibinfo {author}
  {\bibfnamefont {E.}~\bibnamefont {Demler}},\ }\href {\doibase
  10.1103/PhysRevB.96.014512} {\bibfield  {journal} {\bibinfo  {journal} {Phys.
  Rev. B}\ }\textbf {\bibinfo {volume} {96}},\ \bibinfo {pages} {014512}
  (\bibinfo {year} {2017})}\BibitemShut {NoStop}%
\bibitem [{\citenamefont {Murakami}\ \emph {et~al.}(2017)\citenamefont
  {Murakami}, \citenamefont {Tsuji}, \citenamefont {Eckstein},\ and\
  \citenamefont {Werner}}]{murakami2017}%
  \BibitemOpen
  \bibfield  {author} {\bibinfo {author} {\bibfnamefont {Y.}~\bibnamefont
  {Murakami}}, \bibinfo {author} {\bibfnamefont {N.}~\bibnamefont {Tsuji}},
  \bibinfo {author} {\bibfnamefont {M.}~\bibnamefont {Eckstein}}, \ and\
  \bibinfo {author} {\bibfnamefont {P.}~\bibnamefont {Werner}},\ }\href
  {\doibase 10.1103/PhysRevB.96.045125} {\bibfield  {journal} {\bibinfo
  {journal} {Phys. Rev. B}\ }\textbf {\bibinfo {volume} {96}},\ \bibinfo
  {pages} {045125} (\bibinfo {year} {2017})}\BibitemShut {NoStop}%
\bibitem [{\citenamefont {Kennes}\ \emph {et~al.}(2017)\citenamefont {Kennes},
  \citenamefont {Wilner}, \citenamefont {Reichman},\ and\ \citenamefont
  {Millis}}]{kennes2017}%
  \BibitemOpen
  \bibfield  {author} {\bibinfo {author} {\bibfnamefont {D.~M.}\ \bibnamefont
  {Kennes}}, \bibinfo {author} {\bibfnamefont {E.~Y.}\ \bibnamefont {Wilner}},
  \bibinfo {author} {\bibfnamefont {D.~R.}\ \bibnamefont {Reichman}}, \ and\
  \bibinfo {author} {\bibfnamefont {A.~J.}\ \bibnamefont {Millis}},\
  }\href@noop {} {\bibfield  {journal} {\bibinfo  {journal} {Nature Physics}\
  }\textbf {\bibinfo {volume} {13}},\ \bibinfo {pages} {479} (\bibinfo {year}
  {2017})}\BibitemShut {NoStop}%
\bibitem [{\citenamefont {Sentef}(2017)}]{sentef2017}%
  \BibitemOpen
  \bibfield  {author} {\bibinfo {author} {\bibfnamefont {M.~A.}\ \bibnamefont
  {Sentef}},\ }\href {\doibase 10.1103/PhysRevB.95.205111} {\bibfield
  {journal} {\bibinfo  {journal} {Phys. Rev. B}\ }\textbf {\bibinfo {volume}
  {95}},\ \bibinfo {pages} {205111} (\bibinfo {year} {2017})}\BibitemShut
  {NoStop}%
\bibitem [{\citenamefont {Werner}\ and\ \citenamefont
  {Eckstein}(2013)}]{Werner2013}%
  \BibitemOpen
  \bibfield  {author} {\bibinfo {author} {\bibfnamefont {P.}~\bibnamefont
  {Werner}}\ and\ \bibinfo {author} {\bibfnamefont {M.}~\bibnamefont
  {Eckstein}},\ }\href {\doibase 10.1103/PhysRevB.88.165108} {\bibfield
  {journal} {\bibinfo  {journal} {Phys. Rev. B}\ }\textbf {\bibinfo {volume}
  {88}},\ \bibinfo {pages} {165108} (\bibinfo {year} {2013})}\BibitemShut
  {NoStop}%
\bibitem [{\citenamefont {Murakami}\ \emph {et~al.}(2015)\citenamefont
  {Murakami}, \citenamefont {Werner}, \citenamefont {Tsuji},\ and\
  \citenamefont {Aoki}}]{Murakami2015}%
  \BibitemOpen
  \bibfield  {author} {\bibinfo {author} {\bibfnamefont {Y.}~\bibnamefont
  {Murakami}}, \bibinfo {author} {\bibfnamefont {P.}~\bibnamefont {Werner}},
  \bibinfo {author} {\bibfnamefont {N.}~\bibnamefont {Tsuji}}, \ and\ \bibinfo
  {author} {\bibfnamefont {H.}~\bibnamefont {Aoki}},\ }\href {\doibase
  10.1103/PhysRevB.91.045128} {\bibfield  {journal} {\bibinfo  {journal} {Phys.
  Rev. B}\ }\textbf {\bibinfo {volume} {91}},\ \bibinfo {pages} {045128}
  (\bibinfo {year} {2015})}\BibitemShut {NoStop}%
\bibitem [{\citenamefont {Sch\"uler}\ \emph {et~al.}(2016)\citenamefont
  {Sch\"uler}, \citenamefont {Berakdar},\ and\ \citenamefont
  {Pavlyukh}}]{schuler2016}%
  \BibitemOpen
  \bibfield  {author} {\bibinfo {author} {\bibfnamefont {M.}~\bibnamefont
  {Sch\"uler}}, \bibinfo {author} {\bibfnamefont {J.}~\bibnamefont {Berakdar}},
  \ and\ \bibinfo {author} {\bibfnamefont {Y.}~\bibnamefont {Pavlyukh}},\
  }\href {\doibase 10.1103/PhysRevB.93.054303} {\bibfield  {journal} {\bibinfo
  {journal} {Phys. Rev. B}\ }\textbf {\bibinfo {volume} {93}},\ \bibinfo
  {pages} {054303} (\bibinfo {year} {2016})}\BibitemShut {NoStop}%
\bibitem [{\citenamefont {Sentef}\ \emph {et~al.}(2013)\citenamefont {Sentef},
  \citenamefont {Kemper}, \citenamefont {Moritz}, \citenamefont {Freericks},
  \citenamefont {Shen},\ and\ \citenamefont {Devereaux}}]{sentef2013}%
  \BibitemOpen
  \bibfield  {author} {\bibinfo {author} {\bibfnamefont {M.}~\bibnamefont
  {Sentef}}, \bibinfo {author} {\bibfnamefont {A.~F.}\ \bibnamefont {Kemper}},
  \bibinfo {author} {\bibfnamefont {B.}~\bibnamefont {Moritz}}, \bibinfo
  {author} {\bibfnamefont {J.~K.}\ \bibnamefont {Freericks}}, \bibinfo {author}
  {\bibfnamefont {Z.-X.}\ \bibnamefont {Shen}}, \ and\ \bibinfo {author}
  {\bibfnamefont {T.~P.}\ \bibnamefont {Devereaux}},\ }\href {\doibase
  10.1103/PhysRevX.3.041033} {\bibfield  {journal} {\bibinfo  {journal} {Phys.
  Rev. X}\ }\textbf {\bibinfo {volume} {3}},\ \bibinfo {pages} {041033}
  (\bibinfo {year} {2013})}\BibitemShut {NoStop}%
\bibitem [{\citenamefont {Rameau}\ \emph {et~al.}(2016)\citenamefont {Rameau},
  \citenamefont {Freutel}, \citenamefont {Kemper}, \citenamefont {Sentef},
  \citenamefont {Freericks}, \citenamefont {Avigo}, \citenamefont {Ligges},
  \citenamefont {Rettig}, \citenamefont {Yoshida}, \citenamefont {Eisaki} \emph
  {et~al.}}]{rameau2016}%
  \BibitemOpen
  \bibfield  {author} {\bibinfo {author} {\bibfnamefont {J.}~\bibnamefont
  {Rameau}}, \bibinfo {author} {\bibfnamefont {S.}~\bibnamefont {Freutel}},
  \bibinfo {author} {\bibfnamefont {A.}~\bibnamefont {Kemper}}, \bibinfo
  {author} {\bibfnamefont {M.~A.}\ \bibnamefont {Sentef}}, \bibinfo {author}
  {\bibfnamefont {J.}~\bibnamefont {Freericks}}, \bibinfo {author}
  {\bibfnamefont {I.}~\bibnamefont {Avigo}}, \bibinfo {author} {\bibfnamefont
  {M.}~\bibnamefont {Ligges}}, \bibinfo {author} {\bibfnamefont
  {L.}~\bibnamefont {Rettig}}, \bibinfo {author} {\bibfnamefont
  {Y.}~\bibnamefont {Yoshida}}, \bibinfo {author} {\bibfnamefont
  {H.}~\bibnamefont {Eisaki}},  \emph {et~al.},\ }\href@noop {} {\bibfield
  {journal} {\bibinfo  {journal} {Nature communications}\ }\textbf {\bibinfo
  {volume} {7}},\ \bibinfo {pages} {13761} (\bibinfo {year}
  {2016})}\BibitemShut {NoStop}%
\bibitem [{\citenamefont {Kemper}\ \emph {et~al.}(2013)\citenamefont {Kemper},
  \citenamefont {Sentef}, \citenamefont {Moritz}, \citenamefont {Kao},
  \citenamefont {Shen}, \citenamefont {Freericks},\ and\ \citenamefont
  {Devereaux}}]{kemper2013}%
  \BibitemOpen
  \bibfield  {author} {\bibinfo {author} {\bibfnamefont {A.~F.}\ \bibnamefont
  {Kemper}}, \bibinfo {author} {\bibfnamefont {M.}~\bibnamefont {Sentef}},
  \bibinfo {author} {\bibfnamefont {B.}~\bibnamefont {Moritz}}, \bibinfo
  {author} {\bibfnamefont {C.~C.}\ \bibnamefont {Kao}}, \bibinfo {author}
  {\bibfnamefont {Z.~X.}\ \bibnamefont {Shen}}, \bibinfo {author}
  {\bibfnamefont {J.~K.}\ \bibnamefont {Freericks}}, \ and\ \bibinfo {author}
  {\bibfnamefont {T.~P.}\ \bibnamefont {Devereaux}},\ }\href {\doibase
  10.1103/PhysRevB.87.235139} {\bibfield  {journal} {\bibinfo  {journal} {Phys.
  Rev. B}\ }\textbf {\bibinfo {volume} {87}},\ \bibinfo {pages} {235139}
  (\bibinfo {year} {2013})}\BibitemShut {NoStop}%
\bibitem [{\citenamefont {Werner}\ and\ \citenamefont
  {Eckstein}(2015)}]{werner2015}%
  \BibitemOpen
  \bibfield  {author} {\bibinfo {author} {\bibfnamefont {P.}~\bibnamefont
  {Werner}}\ and\ \bibinfo {author} {\bibfnamefont {M.}~\bibnamefont
  {Eckstein}},\ }\href {http://stacks.iop.org/0295-5075/109/i=3/a=37002}
  {\bibfield  {journal} {\bibinfo  {journal} {EPL (Europhysics Letters)}\
  }\textbf {\bibinfo {volume} {109}},\ \bibinfo {pages} {37002} (\bibinfo
  {year} {2015})}\BibitemShut {NoStop}%
\bibitem [{\citenamefont {Grewe}\ and\ \citenamefont
  {Keiter}(1981)}]{grewe1981}%
  \BibitemOpen
  \bibfield  {author} {\bibinfo {author} {\bibfnamefont {N.}~\bibnamefont
  {Grewe}}\ and\ \bibinfo {author} {\bibfnamefont {H.}~\bibnamefont {Keiter}},\
  }\href {\doibase 10.1103/PhysRevB.24.4420} {\bibfield  {journal} {\bibinfo
  {journal} {Phys. Rev. B}\ }\textbf {\bibinfo {volume} {24}},\ \bibinfo
  {pages} {4420} (\bibinfo {year} {1981})}\BibitemShut {NoStop}%
\bibitem [{\citenamefont {Coleman}(1984)}]{coleman1984}%
  \BibitemOpen
  \bibfield  {author} {\bibinfo {author} {\bibfnamefont {P.}~\bibnamefont
  {Coleman}},\ }\href {\doibase 10.1103/PhysRevB.29.3035} {\bibfield  {journal}
  {\bibinfo  {journal} {Phys. Rev. B}\ }\textbf {\bibinfo {volume} {29}},\
  \bibinfo {pages} {3035} (\bibinfo {year} {1984})}\BibitemShut {NoStop}%
\bibitem [{\citenamefont {Eckstein}\ and\ \citenamefont
  {Werner}(2010)}]{eckstein2010}%
  \BibitemOpen
  \bibfield  {author} {\bibinfo {author} {\bibfnamefont {M.}~\bibnamefont
  {Eckstein}}\ and\ \bibinfo {author} {\bibfnamefont {P.}~\bibnamefont
  {Werner}},\ }\href {\doibase 10.1103/PhysRevB.82.115115} {\bibfield
  {journal} {\bibinfo  {journal} {Phys. Rev. B}\ }\textbf {\bibinfo {volume}
  {82}},\ \bibinfo {pages} {115115} (\bibinfo {year} {2010})}\BibitemShut
  {NoStop}%
\bibitem [{\citenamefont {Florens}\ and\ \citenamefont
  {Georges}(2002)}]{Florens2002}%
  \BibitemOpen
  \bibfield  {author} {\bibinfo {author} {\bibfnamefont {S.}~\bibnamefont
  {Florens}}\ and\ \bibinfo {author} {\bibfnamefont {A.}~\bibnamefont
  {Georges}},\ }\href {\doibase 10.1103/PhysRevB.66.165111} {\bibfield
  {journal} {\bibinfo  {journal} {Phys. Rev. B}\ }\textbf {\bibinfo {volume}
  {66}},\ \bibinfo {pages} {165111} (\bibinfo {year} {2002})}\BibitemShut
  {NoStop}%
\bibitem [{\citenamefont {Wilson}(1975)}]{wilson1975}%
  \BibitemOpen
  \bibfield  {author} {\bibinfo {author} {\bibfnamefont {K.~G.}\ \bibnamefont
  {Wilson}},\ }\href@noop {} {\bibfield  {journal} {\bibinfo  {journal} {Rev.
  Mod. Phys.}\ }\textbf {\bibinfo {volume} {47}},\ \bibinfo {pages} {773}
  (\bibinfo {year} {1975})}\BibitemShut {NoStop}%
\bibitem [{\citenamefont {Krishna-murthy}\ \emph {et~al.}(1980)\citenamefont
  {Krishna-murthy}, \citenamefont {Wilkins},\ and\ \citenamefont
  {Wilson}}]{krishna1980a}%
  \BibitemOpen
  \bibfield  {author} {\bibinfo {author} {\bibfnamefont {H.~R.}\ \bibnamefont
  {Krishna-murthy}}, \bibinfo {author} {\bibfnamefont {J.~W.}\ \bibnamefont
  {Wilkins}}, \ and\ \bibinfo {author} {\bibfnamefont {K.~G.}\ \bibnamefont
  {Wilson}},\ }\href@noop {} {\bibfield  {journal} {\bibinfo  {journal} {Phys.
  Rev. B}\ }\textbf {\bibinfo {volume} {21}},\ \bibinfo {pages} {1003}
  (\bibinfo {year} {1980})}\BibitemShut {NoStop}%
\bibitem [{\citenamefont {Bulla}\ \emph {et~al.}(2008)\citenamefont {Bulla},
  \citenamefont {Costi},\ and\ \citenamefont {Pruschke}}]{bulla2008}%
  \BibitemOpen
  \bibfield  {author} {\bibinfo {author} {\bibfnamefont {R.}~\bibnamefont
  {Bulla}}, \bibinfo {author} {\bibfnamefont {T.~A.}\ \bibnamefont {Costi}}, \
  and\ \bibinfo {author} {\bibfnamefont {T.}~\bibnamefont {Pruschke}},\ }\href
  {\doibase 10.1103/RevModPhys.80.395} {\bibfield  {journal} {\bibinfo
  {journal} {Rev. Mod. Phys.}\ }\textbf {\bibinfo {volume} {80}},\ \bibinfo
  {pages} {395} (\bibinfo {year} {2008})}\BibitemShut {NoStop}%
\bibitem [{\citenamefont {Murakami}\ \emph {et~al.}(2013)\citenamefont
  {Murakami}, \citenamefont {Werner}, \citenamefont {Tsuji},\ and\
  \citenamefont {Aoki}}]{murakami2013}%
  \BibitemOpen
  \bibfield  {author} {\bibinfo {author} {\bibfnamefont {Y.}~\bibnamefont
  {Murakami}}, \bibinfo {author} {\bibfnamefont {P.}~\bibnamefont {Werner}},
  \bibinfo {author} {\bibfnamefont {N.}~\bibnamefont {Tsuji}}, \ and\ \bibinfo
  {author} {\bibfnamefont {H.}~\bibnamefont {Aoki}},\ }\href {\doibase
  10.1103/PhysRevB.88.125126} {\bibfield  {journal} {\bibinfo  {journal} {Phys.
  Rev. B}\ }\textbf {\bibinfo {volume} {88}},\ \bibinfo {pages} {125126}
  (\bibinfo {year} {2013})}\BibitemShut {NoStop}%
\bibitem [{\citenamefont {Murakami}\ \emph {et~al.}(2014)\citenamefont
  {Murakami}, \citenamefont {Werner}, \citenamefont {Tsuji},\ and\
  \citenamefont {Aoki}}]{murakami2014}%
  \BibitemOpen
  \bibfield  {author} {\bibinfo {author} {\bibfnamefont {Y.}~\bibnamefont
  {Murakami}}, \bibinfo {author} {\bibfnamefont {P.}~\bibnamefont {Werner}},
  \bibinfo {author} {\bibfnamefont {N.}~\bibnamefont {Tsuji}}, \ and\ \bibinfo
  {author} {\bibfnamefont {H.}~\bibnamefont {Aoki}},\ }\href {\doibase
  10.1103/PhysRevLett.113.266404} {\bibfield  {journal} {\bibinfo  {journal}
  {Phys. Rev. Lett.}\ }\textbf {\bibinfo {volume} {113}},\ \bibinfo {pages}
  {266404} (\bibinfo {year} {2014})}\BibitemShut {NoStop}%
\bibitem [{\citenamefont {Georges}\ \emph {et~al.}(1996)\citenamefont
  {Georges}, \citenamefont {Kotliar}, \citenamefont {Krauth},\ and\
  \citenamefont {Rozenberg}}]{georges1996}%
  \BibitemOpen
  \bibfield  {author} {\bibinfo {author} {\bibfnamefont {A.}~\bibnamefont
  {Georges}}, \bibinfo {author} {\bibfnamefont {G.}~\bibnamefont {Kotliar}},
  \bibinfo {author} {\bibfnamefont {W.}~\bibnamefont {Krauth}}, \ and\ \bibinfo
  {author} {\bibfnamefont {M.~J.}\ \bibnamefont {Rozenberg}},\ }\href {\doibase
  10.1103/RevModPhys.68.13} {\bibfield  {journal} {\bibinfo  {journal} {Rev.
  Mod. Phys.}\ }\textbf {\bibinfo {volume} {68}},\ \bibinfo {pages} {13}
  (\bibinfo {year} {1996})}\BibitemShut {NoStop}%
\bibitem [{\citenamefont {Aoki}\ \emph {et~al.}(2014)\citenamefont {Aoki},
  \citenamefont {Tsuji}, \citenamefont {Eckstein}, \citenamefont {Kollar},
  \citenamefont {Oka},\ and\ \citenamefont {Werner}}]{Aoki2014}%
  \BibitemOpen
  \bibfield  {author} {\bibinfo {author} {\bibfnamefont {H.}~\bibnamefont
  {Aoki}}, \bibinfo {author} {\bibfnamefont {N.}~\bibnamefont {Tsuji}},
  \bibinfo {author} {\bibfnamefont {M.}~\bibnamefont {Eckstein}}, \bibinfo
  {author} {\bibfnamefont {M.}~\bibnamefont {Kollar}}, \bibinfo {author}
  {\bibfnamefont {T.}~\bibnamefont {Oka}}, \ and\ \bibinfo {author}
  {\bibfnamefont {P.}~\bibnamefont {Werner}},\ }\href {\doibase
  10.1103/RevModPhys.86.779} {\bibfield  {journal} {\bibinfo  {journal} {Rev.
  Mod. Phys.}\ }\textbf {\bibinfo {volume} {86}},\ \bibinfo {pages} {779}
  (\bibinfo {year} {2014})}\BibitemShut {NoStop}%
\bibitem [{\citenamefont {Assaad}\ and\ \citenamefont
  {Lang}(2007)}]{Assaad2007}%
  \BibitemOpen
  \bibfield  {author} {\bibinfo {author} {\bibfnamefont {F.~F.}\ \bibnamefont
  {Assaad}}\ and\ \bibinfo {author} {\bibfnamefont {T.~C.}\ \bibnamefont
  {Lang}},\ }\href {\doibase 10.1103/PhysRevB.76.035116} {\bibfield  {journal}
  {\bibinfo  {journal} {Phys. Rev. B}\ }\textbf {\bibinfo {volume} {76}},\
  \bibinfo {pages} {035116} (\bibinfo {year} {2007})}\BibitemShut {NoStop}%
\bibitem [{\citenamefont {Haule}\ \emph {et~al.}(2003)\citenamefont {Haule},
  \citenamefont {Rosch}, \citenamefont {Kroha},\ and\ \citenamefont
  {W{\"{o}}lfle}}]{haule2003}%
  \BibitemOpen
  \bibfield  {author} {\bibinfo {author} {\bibfnamefont {K.}~\bibnamefont
  {Haule}}, \bibinfo {author} {\bibfnamefont {A.}~\bibnamefont {Rosch}},
  \bibinfo {author} {\bibfnamefont {J.}~\bibnamefont {Kroha}}, \ and\ \bibinfo
  {author} {\bibfnamefont {P.}~\bibnamefont {W{\"{o}}lfle}},\ }\href {\doibase
  10.1103/PhysRevB.68.155119} {\bibfield  {journal} {\bibinfo  {journal} {Phys.
  Rev. B}\ }\textbf {\bibinfo {volume} {68}},\ \bibinfo {pages} {155119}
  (\bibinfo {year} {2003})}\BibitemShut {NoStop}%
\bibitem [{\citenamefont {Chen}\ \emph {et~al.}(2016)\citenamefont {Chen},
  \citenamefont {Cohen}, \citenamefont {Millis},\ and\ \citenamefont
  {Reichman}}]{chen2016}%
  \BibitemOpen
  \bibfield  {author} {\bibinfo {author} {\bibfnamefont {H.-T.}\ \bibnamefont
  {Chen}}, \bibinfo {author} {\bibfnamefont {G.}~\bibnamefont {Cohen}},
  \bibinfo {author} {\bibfnamefont {A.~J.}\ \bibnamefont {Millis}}, \ and\
  \bibinfo {author} {\bibfnamefont {D.~R.}\ \bibnamefont {Reichman}},\ }\href
  {\doibase 10.1103/PhysRevB.93.174309} {\bibfield  {journal} {\bibinfo
  {journal} {Phys. Rev. B}\ }\textbf {\bibinfo {volume} {93}},\ \bibinfo
  {pages} {174309} (\bibinfo {year} {2016})}\BibitemShut {NoStop}%
\bibitem [{\citenamefont {{Lang}}\ and\ \citenamefont
  {{Firsov}}(1963)}]{Lang:1963aa}%
  \BibitemOpen
  \bibfield  {author} {\bibinfo {author} {\bibfnamefont {I.~G.}\ \bibnamefont
  {{Lang}}}\ and\ \bibinfo {author} {\bibfnamefont {Y.~A.}\ \bibnamefont
  {{Firsov}}},\ }\href@noop {} {\bibfield  {journal} {\bibinfo  {journal} {Sov.
  Phys. JETP}\ }\textbf {\bibinfo {volume} {16}},\ \bibinfo {pages} {1301}
  (\bibinfo {year} {1963})}\BibitemShut {NoStop}%
\bibitem [{\citenamefont {Hewson}\ and\ \citenamefont
  {Meyer}(2002)}]{hewson2002}%
  \BibitemOpen
  \bibfield  {author} {\bibinfo {author} {\bibfnamefont {A.~C.}\ \bibnamefont
  {Hewson}}\ and\ \bibinfo {author} {\bibfnamefont {D.}~\bibnamefont {Meyer}},\
  }\href@noop {} {\bibfield  {journal} {\bibinfo  {journal} {J. Phys. -
  Condens. Mat.}\ }\textbf {\bibinfo {volume} {14}},\ \bibinfo {pages} {427}
  (\bibinfo {year} {2002})}\BibitemShut {NoStop}%
\bibitem [{\citenamefont {Meyer}\ \emph {et~al.}(2002)\citenamefont {Meyer},
  \citenamefont {Hewson},\ and\ \citenamefont {Bulla}}]{meyer2002}%
  \BibitemOpen
  \bibfield  {author} {\bibinfo {author} {\bibfnamefont {D.}~\bibnamefont
  {Meyer}}, \bibinfo {author} {\bibfnamefont {A.~C.}\ \bibnamefont {Hewson}}, \
  and\ \bibinfo {author} {\bibfnamefont {R.}~\bibnamefont {Bulla}},\
  }\href@noop {} {\bibfield  {journal} {\bibinfo  {journal} {Phys. Rev. Lett.}\
  }\textbf {\bibinfo {volume} {89}},\ \bibinfo {pages} {196401} (\bibinfo
  {year} {2002})}\BibitemShut {NoStop}%
\bibitem [{\citenamefont {Jeon}\ \emph {et~al.}(2003)\citenamefont {Jeon},
  \citenamefont {Park},\ and\ \citenamefont {Choi}}]{jeon2003}%
  \BibitemOpen
  \bibfield  {author} {\bibinfo {author} {\bibfnamefont {G.~S.}\ \bibnamefont
  {Jeon}}, \bibinfo {author} {\bibfnamefont {T.-H.}\ \bibnamefont {Park}}, \
  and\ \bibinfo {author} {\bibfnamefont {H.-Y.}\ \bibnamefont {Choi}},\
  }\href@noop {} {\bibfield  {journal} {\bibinfo  {journal} {Phys. Rev. B}\
  }\textbf {\bibinfo {volume} {68}},\ \bibinfo {pages} {045106} (\bibinfo
  {year} {2003})}\BibitemShut {NoStop}%
\bibitem [{\citenamefont {Cornaglia}\ \emph {et~al.}(2004)\citenamefont
  {Cornaglia}, \citenamefont {Ness},\ and\ \citenamefont
  {Grempel}}]{cornaglia2004}%
  \BibitemOpen
  \bibfield  {author} {\bibinfo {author} {\bibfnamefont {P.~S.}\ \bibnamefont
  {Cornaglia}}, \bibinfo {author} {\bibfnamefont {H.}~\bibnamefont {Ness}}, \
  and\ \bibinfo {author} {\bibfnamefont {D.~R.}\ \bibnamefont {Grempel}},\
  }\href@noop {} {\bibfield  {journal} {\bibinfo  {journal} {Phys. Rev. Lett.}\
  }\textbf {\bibinfo {volume} {93}},\ \bibinfo {pages} {147201} (\bibinfo
  {year} {2004})}\BibitemShut {NoStop}%
\bibitem [{\citenamefont {Koller}\ \emph
  {et~al.}(2004{\natexlab{b}})\citenamefont {Koller}, \citenamefont {Meyer},\
  and\ \citenamefont {Hewson}}]{Koller:2004ic}%
  \BibitemOpen
  \bibfield  {author} {\bibinfo {author} {\bibfnamefont {W.}~\bibnamefont
  {Koller}}, \bibinfo {author} {\bibfnamefont {D.}~\bibnamefont {Meyer}}, \
  and\ \bibinfo {author} {\bibfnamefont {A.}~\bibnamefont {Hewson}},\
  }\href@noop {} {\bibfield  {journal} {\bibinfo  {journal} {Physical Review
  B}\ }\textbf {\bibinfo {volume} {70}},\ \bibinfo {pages} {155103} (\bibinfo
  {year} {2004}{\natexlab{b}})}\BibitemShut {NoStop}%
\bibitem [{\citenamefont {Koller}\ \emph
  {et~al.}(2004{\natexlab{c}})\citenamefont {Koller}, \citenamefont {Meyer},
  \citenamefont {Ono},\ and\ \citenamefont {Hewson}}]{koller2004euro}%
  \BibitemOpen
  \bibfield  {author} {\bibinfo {author} {\bibfnamefont {W.}~\bibnamefont
  {Koller}}, \bibinfo {author} {\bibfnamefont {D.}~\bibnamefont {Meyer}},
  \bibinfo {author} {\bibfnamefont {Y.}~\bibnamefont {Ono}}, \ and\ \bibinfo
  {author} {\bibfnamefont {A.~C.}\ \bibnamefont {Hewson}},\ }\href@noop {}
  {\bibfield  {journal} {\bibinfo  {journal} {Europhys. Lett.}\ }\textbf
  {\bibinfo {volume} {66}},\ \bibinfo {pages} {559} (\bibinfo {year}
  {2004}{\natexlab{c}})}\BibitemShut {NoStop}%
\bibitem [{\citenamefont {Cornaglia}\ \emph {et~al.}(2005)\citenamefont
  {Cornaglia}, \citenamefont {Grempel},\ and\ \citenamefont
  {Ness}}]{cornaglia2005}%
  \BibitemOpen
  \bibfield  {author} {\bibinfo {author} {\bibfnamefont {P.~S.}\ \bibnamefont
  {Cornaglia}}, \bibinfo {author} {\bibfnamefont {D.~R.}\ \bibnamefont
  {Grempel}}, \ and\ \bibinfo {author} {\bibfnamefont {H.}~\bibnamefont
  {Ness}},\ }\href@noop {} {\bibfield  {journal} {\bibinfo  {journal} {Phys.
  Rev. B}\ }\textbf {\bibinfo {volume} {71}},\ \bibinfo {pages} {075320}
  (\bibinfo {year} {2005})}\BibitemShut {NoStop}%
\bibitem [{\citenamefont {Koller}\ \emph {et~al.}(2005)\citenamefont {Koller},
  \citenamefont {Hewson},\ and\ \citenamefont {Edwards}}]{koller2005phonon}%
  \BibitemOpen
  \bibfield  {author} {\bibinfo {author} {\bibfnamefont {W.}~\bibnamefont
  {Koller}}, \bibinfo {author} {\bibfnamefont {A.~C.}\ \bibnamefont {Hewson}},
  \ and\ \bibinfo {author} {\bibfnamefont {D.~M.}\ \bibnamefont {Edwards}},\
  }\href@noop {} {\bibfield  {journal} {\bibinfo  {journal} {Phys. Rev. Lett.}\
  }\textbf {\bibinfo {volume} {95}},\ \bibinfo {pages} {256401} (\bibinfo
  {year} {2005})}\BibitemShut {NoStop}%
\bibitem [{\citenamefont {Cornaglia}\ \emph {et~al.}(2007)\citenamefont
  {Cornaglia}, \citenamefont {Usaj},\ and\ \citenamefont
  {Balseiro}}]{cornaglia2007}%
  \BibitemOpen
  \bibfield  {author} {\bibinfo {author} {\bibfnamefont {P.~S.}\ \bibnamefont
  {Cornaglia}}, \bibinfo {author} {\bibfnamefont {G.}~\bibnamefont {Usaj}}, \
  and\ \bibinfo {author} {\bibfnamefont {C.~A.}\ \bibnamefont {Balseiro}},\
  }\href@noop {} {\bibfield  {journal} {\bibinfo  {journal} {Phys. Rev. B}\
  }\textbf {\bibinfo {volume} {76}},\ \bibinfo {pages} {241403(R)} (\bibinfo
  {year} {2007})}\BibitemShut {NoStop}%
\bibitem [{\citenamefont {Bauer}(2010)}]{Bauer:2010hx}%
  \BibitemOpen
  \bibfield  {author} {\bibinfo {author} {\bibfnamefont {J.}~\bibnamefont
  {Bauer}},\ }\href@noop {} {\bibfield  {journal} {\bibinfo  {journal}
  {Europhysics Letters}\ }\textbf {\bibinfo {volume} {90}},\ \bibinfo {pages}
  {27002} (\bibinfo {year} {2010})}\BibitemShut {NoStop}%
\bibitem [{\citenamefont {Bauer}\ and\ \citenamefont
  {Hewson}(2010)}]{Bauer:2010by}%
  \BibitemOpen
  \bibfield  {author} {\bibinfo {author} {\bibfnamefont {J.}~\bibnamefont
  {Bauer}}\ and\ \bibinfo {author} {\bibfnamefont {A.~C.}\ \bibnamefont
  {Hewson}},\ }\href@noop {} {\bibfield  {journal} {\bibinfo  {journal}
  {Physical Review B}\ }\textbf {\bibinfo {volume} {81}},\ \bibinfo {pages}
  {235113} (\bibinfo {year} {2010})}\BibitemShut {NoStop}%
\bibitem [{\citenamefont {Gole\v{z}}\ \emph {et~al.}(2012)\citenamefont
  {Gole\v{z}}, \citenamefont {Bon\v{c}a},\ and\ \citenamefont
  {\v{Z}itko}}]{golez2012}%
  \BibitemOpen
  \bibfield  {author} {\bibinfo {author} {\bibfnamefont {D.}~\bibnamefont
  {Gole\v{z}}}, \bibinfo {author} {\bibfnamefont {J.}~\bibnamefont
  {Bon\v{c}a}}, \ and\ \bibinfo {author} {\bibfnamefont {R.}~\bibnamefont
  {\v{Z}itko}},\ }\href@noop {} {\bibfield  {journal} {\bibinfo  {journal}
  {Phys. Rev. B}\ }\textbf {\bibinfo {volume} {86}},\ \bibinfo {pages} {085142}
  (\bibinfo {year} {2012})}\BibitemShut {NoStop}%
\bibitem [{\citenamefont {\v{Z}itko}\ and\ \citenamefont
  {Pruschke}(2009)}]{resolution}%
  \BibitemOpen
  \bibfield  {author} {\bibinfo {author} {\bibfnamefont {R.}~\bibnamefont
  {\v{Z}itko}}\ and\ \bibinfo {author} {\bibfnamefont {T.}~\bibnamefont
  {Pruschke}},\ }\href@noop {} {\bibfield  {journal} {\bibinfo  {journal}
  {Phys. Rev. B}\ }\textbf {\bibinfo {volume} {79}},\ \bibinfo {pages} {085106}
  (\bibinfo {year} {2009})}\BibitemShut {NoStop}%
\bibitem [{\citenamefont {\v{Z}itko}(2009)}]{odesolv}%
  \BibitemOpen
  \bibfield  {author} {\bibinfo {author} {\bibfnamefont {R.}~\bibnamefont
  {\v{Z}itko}},\ }\href@noop {} {\bibfield  {journal} {\bibinfo  {journal}
  {Comp. Phys. Comm.}\ }\textbf {\bibinfo {volume} {180}},\ \bibinfo {pages}
  {1271} (\bibinfo {year} {2009})}\BibitemShut {NoStop}%
\bibitem [{\citenamefont {Anders}\ and\ \citenamefont
  {Schiller}(2005)}]{anders2005}%
  \BibitemOpen
  \bibfield  {author} {\bibinfo {author} {\bibfnamefont {F.~B.}\ \bibnamefont
  {Anders}}\ and\ \bibinfo {author} {\bibfnamefont {A.}~\bibnamefont
  {Schiller}},\ }\href@noop {} {\bibfield  {journal} {\bibinfo  {journal}
  {Phys. Rev. Lett.}\ }\textbf {\bibinfo {volume} {95}},\ \bibinfo {pages}
  {196801} (\bibinfo {year} {2005})}\BibitemShut {NoStop}%
\bibitem [{\citenamefont {Peters}\ \emph {et~al.}(2006)\citenamefont {Peters},
  \citenamefont {Pruschke},\ and\ \citenamefont {Anders}}]{peters2006}%
  \BibitemOpen
  \bibfield  {author} {\bibinfo {author} {\bibfnamefont {R.}~\bibnamefont
  {Peters}}, \bibinfo {author} {\bibfnamefont {T.}~\bibnamefont {Pruschke}}, \
  and\ \bibinfo {author} {\bibfnamefont {F.~B.}\ \bibnamefont {Anders}},\
  }\href@noop {} {\bibfield  {journal} {\bibinfo  {journal} {Phys. Rev. B}\
  }\textbf {\bibinfo {volume} {74}},\ \bibinfo {pages} {245114} (\bibinfo
  {year} {2006})}\BibitemShut {NoStop}%
\bibitem [{\citenamefont {Weichselbaum}\ and\ \citenamefont {von
  Delft}(2007)}]{weichselbaum2007}%
  \BibitemOpen
  \bibfield  {author} {\bibinfo {author} {\bibfnamefont {A.}~\bibnamefont
  {Weichselbaum}}\ and\ \bibinfo {author} {\bibfnamefont {J.}~\bibnamefont {von
  Delft}},\ }\href@noop {} {\bibfield  {journal} {\bibinfo  {journal} {Phys.
  Rev. Lett.}\ }\textbf {\bibinfo {volume} {99}},\ \bibinfo {pages} {076402}
  (\bibinfo {year} {2007})}\BibitemShut {NoStop}%
\bibitem [{\citenamefont {Mahan}(2000)}]{mahan2000}%
  \BibitemOpen
  \bibfield  {author} {\bibinfo {author} {\bibfnamefont {G.}~\bibnamefont
  {Mahan}},\ }\href@noop {} {\emph {\bibinfo {title} {Many-particle physics}}}\
  (\bibinfo  {publisher} {Springer},\ \bibinfo {year} {2000})\BibitemShut
  {NoStop}%
\bibitem [{\citenamefont {Freericks}\ \emph {et~al.}(2009)\citenamefont
  {Freericks}, \citenamefont {Krishnamurthy},\ and\ \citenamefont
  {Pruschke}}]{freericks09}%
  \BibitemOpen
  \bibfield  {author} {\bibinfo {author} {\bibfnamefont {J.~K.}\ \bibnamefont
  {Freericks}}, \bibinfo {author} {\bibfnamefont {H.~R.}\ \bibnamefont
  {Krishnamurthy}}, \ and\ \bibinfo {author} {\bibfnamefont {T.}~\bibnamefont
  {Pruschke}},\ }\href@noop {} {\bibfield  {journal} {\bibinfo  {journal}
  {Phys. Rev. Lett.}\ }\textbf {\bibinfo {volume} {102}},\ \bibinfo {pages}
  {136401} (\bibinfo {year} {2009})}\BibitemShut {NoStop}%
\bibitem [{\citenamefont {Weber}(1956)}]{Weber1956}%
  \BibitemOpen
  \bibfield  {author} {\bibinfo {author} {\bibfnamefont {J.}~\bibnamefont
  {Weber}},\ }\href {\doibase 10.1103/PhysRev.101.1620} {\bibfield  {journal}
  {\bibinfo  {journal} {Phys. Rev.}\ }\textbf {\bibinfo {volume} {101}},\
  \bibinfo {pages} {1620} (\bibinfo {year} {1956})}\BibitemShut {NoStop}%
\bibitem [{\citenamefont {Iwai}\ \emph {et~al.}(2003)\citenamefont {Iwai},
  \citenamefont {Ono}, \citenamefont {Maeda}, \citenamefont {Matsuzaki},
  \citenamefont {Kishida}, \citenamefont {Okamoto},\ and\ \citenamefont
  {Tokura}}]{Iwai2003}%
  \BibitemOpen
  \bibfield  {author} {\bibinfo {author} {\bibfnamefont {S.}~\bibnamefont
  {Iwai}}, \bibinfo {author} {\bibfnamefont {M.}~\bibnamefont {Ono}}, \bibinfo
  {author} {\bibfnamefont {A.}~\bibnamefont {Maeda}}, \bibinfo {author}
  {\bibfnamefont {H.}~\bibnamefont {Matsuzaki}}, \bibinfo {author}
  {\bibfnamefont {H.}~\bibnamefont {Kishida}}, \bibinfo {author} {\bibfnamefont
  {H.}~\bibnamefont {Okamoto}}, \ and\ \bibinfo {author} {\bibfnamefont
  {Y.}~\bibnamefont {Tokura}},\ }\href {\doibase 10.1103/PhysRevLett.91.057401}
  {\bibfield  {journal} {\bibinfo  {journal} {Phys. Rev. Lett.}\ }\textbf
  {\bibinfo {volume} {91}},\ \bibinfo {pages} {057401} (\bibinfo {year}
  {2003})}\BibitemShut {NoStop}%
\bibitem [{\citenamefont {Lysenko}\ \emph {et~al.}(2007)\citenamefont
  {Lysenko}, \citenamefont {R\'ua}, \citenamefont {Vikhnin}, \citenamefont
  {Fern\'andez},\ and\ \citenamefont {Liu}}]{Lysenko2007}%
  \BibitemOpen
  \bibfield  {author} {\bibinfo {author} {\bibfnamefont {S.}~\bibnamefont
  {Lysenko}}, \bibinfo {author} {\bibfnamefont {A.}~\bibnamefont {R\'ua}},
  \bibinfo {author} {\bibfnamefont {V.}~\bibnamefont {Vikhnin}}, \bibinfo
  {author} {\bibfnamefont {F.}~\bibnamefont {Fern\'andez}}, \ and\ \bibinfo
  {author} {\bibfnamefont {H.}~\bibnamefont {Liu}},\ }\href {\doibase
  10.1103/PhysRevB.76.035104} {\bibfield  {journal} {\bibinfo  {journal} {Phys.
  Rev. B}\ }\textbf {\bibinfo {volume} {76}},\ \bibinfo {pages} {035104}
  (\bibinfo {year} {2007})}\BibitemShut {NoStop}%
\bibitem [{\citenamefont {Demsar}\ \emph {et~al.}(2003)\citenamefont {Demsar},
  \citenamefont {Averitt}, \citenamefont {Ahn}, \citenamefont {Graf},
  \citenamefont {Trugman}, \citenamefont {Kabanov}, \citenamefont {Sarrao},\
  and\ \citenamefont {Taylor}}]{Demsar2003}%
  \BibitemOpen
  \bibfield  {author} {\bibinfo {author} {\bibfnamefont {J.}~\bibnamefont
  {Demsar}}, \bibinfo {author} {\bibfnamefont {R.~D.}\ \bibnamefont {Averitt}},
  \bibinfo {author} {\bibfnamefont {K.~H.}\ \bibnamefont {Ahn}}, \bibinfo
  {author} {\bibfnamefont {M.~J.}\ \bibnamefont {Graf}}, \bibinfo {author}
  {\bibfnamefont {S.~A.}\ \bibnamefont {Trugman}}, \bibinfo {author}
  {\bibfnamefont {V.~V.}\ \bibnamefont {Kabanov}}, \bibinfo {author}
  {\bibfnamefont {J.~L.}\ \bibnamefont {Sarrao}}, \ and\ \bibinfo {author}
  {\bibfnamefont {A.~J.}\ \bibnamefont {Taylor}},\ }\href {\doibase
  10.1103/PhysRevLett.91.027401} {\bibfield  {journal} {\bibinfo  {journal}
  {Phys. Rev. Lett.}\ }\textbf {\bibinfo {volume} {91}},\ \bibinfo {pages}
  {027401} (\bibinfo {year} {2003})}\BibitemShut {NoStop}%
\bibitem [{\citenamefont {Werner}\ and\ \citenamefont
  {Millis}(2006)}]{Werner:2006qy}%
  \BibitemOpen
  \bibfield  {author} {\bibinfo {author} {\bibfnamefont {P.}~\bibnamefont
  {Werner}}\ and\ \bibinfo {author} {\bibfnamefont {A.~J.}\ \bibnamefont
  {Millis}},\ }\href {\doibase 10.1103/PhysRevB.74.155107} {\bibfield
  {journal} {\bibinfo  {journal} {Phys. Rev. B}\ }\textbf {\bibinfo {volume}
  {74}},\ \bibinfo {pages} {155107} (\bibinfo {year} {2006})}\BibitemShut
  {NoStop}%
\end{thebibliography}%

\end{document}